\documentclass[a4paper,11pt]{article}
\usepackage{jheppub} 
\usepackage{lineno}
\usepackage{comment}
\usepackage{bm}
\usepackage{graphicx}
\usepackage{amsmath}
\usepackage{mathrsfs}
\usepackage[super]{nth}
\usepackage{mathbbol}
\usepackage{color}

\usepackage[normalem]{ulem}  

\newcommand{\old}[1]{{\color[rgb]{0.7,0,0.7}\sout{#1}}}

\arxivnumber{} 

\title{Thermodynamics of accelerating AdS$_4$ black holes from the covariant phase space}

\author{Hyojoong Kim,}
\author{Nakwoo Kim,}
\author{Yein Lee}
\author{and Aaron Poole}
\affiliation{Department of Physics and Research Institute of Basic Science, Kyung Hee University,
Seoul 02447, Korea}

\emailAdd{h.kim@khu.ac.kr}
\emailAdd{nkim@khu.ac.kr}
\emailAdd{lyi126@khu.ac.kr}
\emailAdd{apoole@khu.ac.kr}

\abstract{We study the charges and first law of thermodynamics for accelerating, non-rotating black holes with dyonic charges in AdS$_4$ using the covariant phase space formalism. In order to apply the formalism to these solutions (which are asymptotically locally AdS and admit a non-smooth conformal boundary $\mathscr{I}$) we make two key improvements: 1) We relax the requirement to impose Dirichlet boundary conditions and demand merely a well-posed variational problem. 2) We keep careful track of the codimension-2 corner term induced by the holographic counterterms, a necessary requirement due to the presence of ``cosmic strings" piercing $\mathscr{I}$. Using these improvements we are able to match the holographic Noether charges to the Wald Hamiltonians of the covariant phase space and derive the first law of black hole thermodynamics with the correct ``thermodynamic length'' terms arising from the strings. We investigate the relationship between the charges imposed by supersymmetry and show that our first law can be consistently applied to various classes of non-supersymmetric solutions for which the cross-sections of the horizon are spindles.}

\begin{document}
\maketitle
\flushbottom

\section{Introduction} \label{sec:intro}

The understanding of black holes as thermodynamic objects is one of the key directions in uncovering the quantum nature of gravity. The origin of this field lies in the pioneering work by Bekenstein \cite{Bekenstein:1972tm, Bekenstein:1973ur}, who conjectured that the entropy of a black hole should be proportional to the  horizon area $\mathcal{A}$, and later Hawking \cite{Hawking:1974rv}, where it was demonstrated that by taking into account the effects of quantum particle creation near the horizon, black holes possess a temperature $T$. This confirmed Bekenstein's conjecture and resulted in the famous Bekenstein-Hawking entropy formula
\begin{equation} \label{eq: BH_entropy}
S_{\text{BH}} = \frac{\mathcal{A}}{4G}. 
\end{equation}

Alongside this identification of black hole entropy was the realisation that black holes obey certain laws of mechanics closely analogous to the ordinary laws of thermodynamics. Of particular focus in this work will be the \textit{first law} of black hole thermodynamics, which was originally formulated for stationary, asmyptotically flat black holes as \cite{Bardeen:1973gs}
\begin{equation} \label{eq: first_law_af}
    \delta \mathcal{M} = T\delta S_{\text{BH}} + \Omega_{\mathcal{H}} \delta J + \Phi_e \delta Q_e,
\end{equation}
a formula which relates variations in the charges $\mathcal{M}, J, Q_e$ (mass, angular momentum and electric charge) of the black hole to variations in the entropy ($\Omega_{\mathcal{H}}$ is the angular velocity of the horizon and $\Phi_e$ the electrostatic potential). Such a formula was generalised by Wald \cite{Wald:1993nt} to all diffeomorphism invariant theories of gravity, (i.e. beyond just general relativity) with the entropy taking the form of a local integral over the bifurcation surface of the horizon $\Sigma_{\mathcal{H}}$
\begin{equation}
    S_{\text{BH}} = \frac{2\pi}{\kappa_{\text{sg}}} \int_{\Sigma_{\mathcal{H}}} \mathbf{Q},
\end{equation}
where $\mathbf{Q}$ is the so-called Noether charge $(d-2)$-form of the theory and $\kappa_{\text{sg}}$ the surface gravity of the black hole. This approach uses a technique known as the \textit{covariant phase space formalism} \cite{Crnkovic:1986ex, Wald:1993nt, Iyer:1994ys} and not only has the advantage of extending to other theories but also gives an elegant geometrical derivation of the first law in terms of covariant expressions, most importantly the local formula for the entropy above.

In this work we will study \textit{accelerating} black holes in \textit{asymptotically locally anti-de Sitter} (AlAdS) spacetime using the covariant phase space formalism. Black holes in AdS have proven to be particularly rich hunting grounds for those looking to understand their quantum properties thanks to the AdS/CFT correspondence \cite{Maldacena:1997re, Witten:1998qj, Gubser:1998bc}. This allows for entropy counting in the gravitational side to be reformulated in terms of an index computation in the dual CFT, see e.g. \cite{Benini:2015noa, Benini:2015eyy, Benini:2016hjo, Benini:2016rke} 
for black holes in $d=4$ and \cite{Cabo-Bizet:2018ehj, Choi:2018hmj, Benini:2018ywd} for $d=5$. These techniques have made it possible to recover the Bekenstein-Hawking entropy (\ref{eq: BH_entropy}) from the dual theory. For classical AdS gravity, the analogous first law to (\ref{eq: first_law_af}) has been derived for a wide class of AlAdS black holes \cite{Papadimitriou:2005ii} using the covariant phase space \cite{Wald:1993nt, Iyer:1994ys} together with the necessary implementation of \textit{holographic renormalisation} \cite{Henningson:1998gx, Balasubramanian:1999re, deHaro:2000vlm, Skenderis:2000in, Skenderis:2002wp, Papadimitriou:2004ap} at the level of the on-shell action. The use of the covariant phase space has been extended to theories beyond those initially considered in \cite{Papadimitriou:2005ii} (see for example the recent works \cite{Cassani:2022lrk, Awad:2022exw, Cassani:2023vsa} on various $d=5$ supergravity theories) but has not yet been adapted to accelerating AdS$_4$ black holes. This important gap in the literature will be addressed in this work.  

The progenitive accelerating black hole in AdS$_4$ is the famous C-metric solution  \cite{Kinnersley:1970zw}, a member of the more general Plebanski-Demia\'nski class of stationary, axisymmetric solutions \cite{Plebanski:1976gy, Dias:2002mi, Griffiths:2005qp, Podolsky:2022xxd}. These black holes possess conical singularities due to the presence of cosmic strings stretching from the horizon of the black hole out to infinity. The cosmic strings have associated tensions which exert a force on the black hole, resulting in acceleration and displacing the object from the ``centre" of the spacetime. In this work we will consider black holes which are said to be \textit{slowly} accelerating \cite{Podolsky:2002nk, Krtous:2005ej}, meaning that they possess an event horizon but no acceleration horizons. We will take the solutions to have charges corresponding to mass, electric, and magnetic charges but, importantly, not rotation. We will thus work with static solutions with 
\begin{equation}
    J = 0,
\end{equation}
for reasons we will discuss in the main text. As we shall see, the fact that these spacetimes contain conical singularities, together with the fact that one cannot apply Dirichlet boundary conditions when varying all of the parameters are the crucial obstruction in applying the methods of \cite{Papadimitriou:2005ii}. In this work we will provide a suitable extension of the methods developed in \cite{Papadimitriou:2005ii} in order to discuss the charges and thermodynamics of accelerating solutions. 

The covariant phase space \cite{Wald:1993nt, Iyer:1994ys, Papadimitriou:2005ii} has yet to be applied to accelerating AlAdS black holes, although analysing their charges and associated thermodynamics using different techniques have been the study of a slew of recent work \cite{Appels:2016uha, Appels:2017xoe, Gregory:2017ogk, Anabalon:2018ydc, Anabalon:2018qfv, Cassani:2021dwa} which we will follow closely (see also \cite{Astorino:2016xiy, Astorino:2016ybm, Jafarzade:2017kin, EslamPanah:2019szt, EslamPanah:2022ihg} for related works). A major feature present in these papers was the seeming inevitability of being forced to choose a particular parameter-dependent normalisation of the time coordinate in order to arrive at the correct form of the first law. Some justification for this was given in \cite{Anabalon:2018ydc} in terms of asymptotic observers, although the conformal invariance at the boundary should negate the need to study a particular representative of the conformal class. This scaling is thus a somewhat unsatisfactory feature which is also not at all clear from the perspective of the dual field theory. We note that this story is somewhat similar in spirit to that of \cite{Gibbons:2004ai} where it was argued that the normalisation of the Killing vector in the first law was crucial in defining the charges and first law, before \cite{Papadimitriou:2005ii} demonstrated that the correct application of the covariant phase space overrides such issues and the first law is satisfied for all non-accelerating AlAdS black holes. It is in this vein that we expect the application of the covariant phase space formalism to shed new light on the time scaling and uncover the physics of this poorly-understood feature of black hole thermodynamics. In particular, we will show that the previous choice of the time scaling is only a well-posed choice when one also fixes the overall conical deficit in the spacetime. In this work we will consider the more general problem of well-posed variations, without explicitly fixing the time scaling. 

Accelerating solutions are also of interest in the field of \textit{supergravity} due to their relation to the field of \textit{spindle solutions} \cite{Cassani:2021dwa, Ferrero:2020laf, Ferrero:2020twa, Ferrero:2021wvk}.
 If the cosmic strings associated to acceleration are arranged in a particular way, then the surfaces of constant time and radius $\Sigma$ can be given the topology of a spindle $\Sigma \cong \mathbb{WCP}^1_{[n_-, n_+]}$, a complex projective space  parameterised by two coprime positive integers $\{n_-, n_+\}$. Such solutions are interesting because despite exhibiting conical singularities in $d=4$, they are rendered completely smooth in $d=11$ supergravity when uplifted on a suitably chosen Sasaki-Einstein manifold $SE_7$ \cite{Ferrero:2020twa}. Following in the style of \cite{Cassani:2021dwa}, we will work in $d=4$ for the entirety of this paper and the uplift will not come into play. Supersymmetry will be preserved in $d=11$ if it is satisfied in $d=4$ and thus it is of interest to constrain the parameters of the solution via the supersymmetry conditions discussed in \cite{Klemm:2013eca, Ferrero:2020twa, Cassani:2021dwa}. 

A further important subclass of these solutions are the \textit{supersymmetric and extremal} AdS$_4$ black holes with $\Sigma \cong \mathbb{WCP}^1_{[n_-, n_+]}$. These exhibit a near-horizon geometry of AdS$_2 \times \mathbb{WCP}^1_{[n_-, n_+]}$ and uplift in $d=11$ to solutions with near horizon regions of the form AdS$_2 \times Y_9$, where $Y_9$ is a geometry of the type discussed in \cite{Gauntlett:2006ns, Gauntlett:2007ts}. A thorough understanding of the $d=4$ solutions may also shed new light into the class of solutions with an AdS$_2$ factor and thus one is also motivated to apply extremality as well as supersymmetry for solutions in $d=4$.  We will use the supersymmetry relations in order to derive a ``supersymmetric locus" of conserved charges but will stop short of being able to apply our first law to the supersymmetric solutions. This is because such solutions must contain either acceleration horizons (when extremal) or naked singularities (when non-extremal) \cite{Ferrero:2020twa} and thus fall outside the class of slowly accelerating solutions that we consider. Instead, we will apply our law to non-supersymmetric spindles, including the classes of \textit{close-to-supersymmetric} and \textit{close-to-supersymmetric and close-to-extremal} solutions, which are smoothly connected to the supersymmetric cases \cite{Ferrero:2020twa}.

This paper is organised as follows: in Section \ref{sec: Accelerating solutions} we provide a brief introduction to the family of solutions that we consider and discuss the physics of the parameters that specify the solutions. In Section \ref{sec: asymptotic_analysis} we perform a careful asymptotic analysis of the metric and gauge field which specify the solution. This includes a presentation of the Fefferman-Graham \cite{Fefferman:1985zza} expansion for the metric as well as an analysis of the boundary Cotton tensor. We use the asymptotic analysis to discuss the variational problem and derive a master equation for well-posedness. In Section \ref{sec: Charges} we use the covariant phase space formalism \cite{Wald:1993nt, Papadimitriou:2005ii} to construct the conserved charges for the solution. This section includes an introduction to the formalism as well as a discussion of the required corner modifications in order to allow for application to spacetimes with conical singularities. We use this to give expressions for the mass, electric, and magnetic charges of the accelerating solutions. In Section \ref{sec: first_law} we focus on the thermodynamics of accelerating black holes, again using the covariant phase space approach to write down the first law of thermodynamics. We provide a comment on the form of our law relative to others in the literature \cite{Appels:2016uha, Appels:2017xoe, Gregory:2017ogk, Anabalon:2018ydc, Anabalon:2018qfv, Cassani:2021dwa}. In Section \ref{sec: spindles_susy} we provide an application of our results for the conserved charges and first law to spindle solutions: we fix the string tensions and apply various other constraints related to supersymmetry and extremality. In Section \ref{sec: conclusions} we conclude and discuss some interesting directions for future work. Also included are two appendices: Appendix \ref{sec: Magnetic_charges} discusses the nature of magnetic charges from the covariant phase space and Appendix \ref{sec: horizon_poly} provides a detailed comparison with other literature \cite{Appels:2016uha, Appels:2017xoe, Gregory:2017ogk, Anabalon:2018ydc, Anabalon:2018qfv, Cassani:2021dwa}. This includes a demonstration of equivalence between the covariant phase space and ``horizon polynomial" methods of deriving the first law, as well as a more detailed discussion concerning the discrepancies in our laws. 

\section{Accelerating solutions} \label{sec: Accelerating solutions}

In this work we study Einstein-Maxwell theory in the presence of a cosmological constant $\Lambda = -3/\ell^2 < 0$ on a $d=4$ dimensional spacetime manifold $M$. We will consider the following bulk action: 
\begin{equation} \label{eq: bulk_action}
    S_{\text{bulk}} =\frac{1}{16 \pi G} \int_{M} (R -2 \Lambda) \bm{\epsilon} - 2 \mathbf{F} \wedge * \mathbf{F},
\end{equation}
where $\bm{\epsilon}$ is the volume 4-form, oriented such that $\epsilon_{0123} = \sqrt{-g}$. $\mathbf{F} = d \mathbf{A}$ is the 2-form field strength tensor with $\mathbf{A}$ the 1-form gauge potential. We will discuss the possibility of adding a purely topological term related to magnetic charges in appendix \ref{sec: Magnetic_charges} but this bulk action will be sufficient for all of our analysis of the first law. 

We consider the following family of accelerating, static solutions \cite{Kinnersley:1970zw, Plebanski:1976gy, Dias:2002mi, Hong:2003gx, Griffiths:2005qp, Podolsky:2022xxd} with metric
\begin{equation} \label{eq: charged_acc_bh_metric}
    ds^2 = \frac{1}{H^2} \left\{ - \frac{Q}{r^2} \frac{1}{\kappa^2}  dt^2 + \frac{r^2}{Q} dr^2 + \frac{r^2}{P} d\theta^2 + P r^2 K^2 \sin^2 \theta  d\varphi^2 \right\},
\end{equation}
and gauge field 
\begin{equation} \label{eq: charged_acc_bh_gauge_field}
 \mathbf{A} = - \frac{e}{r} \frac{1}{\kappa} dt - g \cos \theta K d\varphi,
\end{equation}
where
\begin{align}
    \begin{split} \label{eq: metric_functions}
     H(r, \theta) & = 1 - \alpha r \cos \theta, \qquad Q(r) = (r^2 -2 m r + e^2 + g^2)(1-\alpha^2 r^2) + \frac{r^4}{\ell^2}, \\ 
        P(\theta) & = 1 - 2\alpha m \cos \theta + \alpha^2 (e^2 + g^2) \cos^2 \theta.
   \end{split}
\end{align}
The solution is determined by five physical parameters: $\{m, e, g, \alpha, K\}$ as well as the cosmological constant $\Lambda=-3/\ell^2$ and the time scaling $\kappa >0$ which as in  \cite{Anabalon:2018qfv, Cassani:2021dwa} we take to be a spacetime constant. The physical parameters have the rough identifications as corresponding to mass, electric charge, magnetic charge, acceleration and deficit angle\footnote{Note that our choice of deficit parameter $K$ is related to that of \cite{Anabalon:2018ydc} by $K \rightarrow 1/K$.}  respectively and thus we will refer to them as such throughout the text. We will make explicit their relation to the true charges of the spacetime in Section \ref{sec: Charges}. Following \cite{Ferrero:2020twa, Cassani:2021dwa} we consider w.l.o.g. the following ranges of parameters
\begin{equation}
m, K > 0, \qquad \alpha, e, g \geq 0,
\end{equation}
although in general we will be interested in the case of all parameters being strictly positive.

The metric is determined by three functions $\{H, Q, P\}$ given in equation (\ref{eq: metric_functions}) which we now describe in some detail in order to explain the physics of this solution. Firstly, $Q$ is the \textit{horizon polynomial} and the roots of $Q$ are the locations of horizons in the spacetime. We will demand that the solution contains a black hole and thus the largest positive root $r_+$ corresponds to the location of the (outer) event horizon $\mathcal{H}$ in the spacetime:
\begin{equation}
    Q(r_+) = 0, \qquad r_+ > 0.
\end{equation}
In the entirety of this work, following \cite{Podolsky:2002nk, Krtous:2005ej, Appels:2016uha, Appels:2017xoe, Gregory:2017ogk, Anabalon:2018ydc, Anabalon:2018qfv, Cassani:2021dwa}, we will restrict to the case of \textit{slowly accelerating} solutions, i.e. those without an acceleration horizon.
This assumption corresponds to there being no further roots of $Q$ between $\mathcal{H}$ and the conformal boundary $\mathscr{I}$. For a technical discussion of this in terms of the parameters of the solution we point the reader to \cite{Podolsky:2002nk, Krtous:2005ej, Gregory:2017ogk}, an analysis which we omit here as we will only use this assumption implicitly. We will be interested in studying the region of the solution outside the black hole, and thus we restrict consideration to the coordinate range  
\begin{equation} \label{eq: r_lower_bound}
r > r_+ > 0.
\end{equation}

$H$ is the \textit{conformal factor} and thus the conformal boundary $\mathscr{I}$ is located at $H=0$. This sets the upper bound on the radial coordinate as  
\begin{equation} \label{eq: r_upper_bound}
 r < \frac{1}{\alpha \cos \theta},
\end{equation}
where we note that $r$ is not a good coordinate to analyse the conformal boundary for $\theta \geq \pi/2$ and we will utilise a different choice for the asymptotic analysis in Section \ref{sec: asymptotic_analysis}. Combining equations (\ref{eq: r_lower_bound}) and (\ref{eq: r_upper_bound}) in the region of validity, we note that this sets 
\begin{equation}
    r_+ < \frac{1}{\alpha},
\end{equation}
a condition which can be physically interpreted as ensuring that the horizon does not touch the conformal boundary \cite{Cassani:2021dwa}.

$P$ is a function which encodes the fact that the spacetime contains conical singularities, physically interpreted as cosmic strings stretching from $\mathcal{H}$ to $\mathscr{I}$. In order to see this explicitly \cite{Anabalon:2018ydc, Ferrero:2020twa}, one can perform an analysis of the metric near the poles of the azimuthal coordinate $\theta_{\pm}$
\begin{equation}
\theta_{-} = 0, \qquad \theta_{+} = \pi,
\end{equation}
which are the North and South poles respectively. Near the poles, the metric on the constant $(t, r)$ surfaces takes the form \cite{Anabalon:2018ydc, Ferrero:2020twa}
\begin{equation} \label{eq: angular_metric}
    ds^2_{\theta, \varphi} \simeq \left[ \frac{r^2}{P H^2} \right]_{\theta = \theta_{\pm}} [d\theta^2 + P_{\pm}^2 K^2 (\theta- \theta_{\pm} )^2 d\varphi^2 ],
\end{equation}
where
\begin{equation}
    P_{\pm} = P(\theta_{\pm}) = \Xi \pm 2 \alpha m,
\end{equation}
and following \cite{Anabalon:2018ydc, Ferrero:2020twa, Cassani:2021dwa} we have introduced 
\begin{equation} \label{eq: Xi}
\Xi = 1 + \alpha^2(e^2 +g^2).
\end{equation}
Returning to (\ref{eq: angular_metric}), we note that $\varphi$ is a $2\pi$-periodic coordinate and thus the metric near each pole takes the a form similar to the usual plane polar coordinates on $\mathbb{R}^2$ with $\theta$ acting a radial coordinate and $\varphi$ the polar angle. In order to remove the possibility of conical singularities, one needs to choose the parameter $K$ s.t. $P_{\pm} K =1$, although this is clearly impossible when $P_- \neq P_+ \iff \alpha m \neq 0$. The resulting spacetime thus contains conical singularities at $\theta_{\pm}$, with deficit angles given by
\begin{equation}
    \delta_{\pm} = 2\pi (1 - P_{\pm} K).
\end{equation}
We note that one can choose $K=1/P_-$ or $K=1/P_+$ in order to remove one of the singularities and leave a spacetime with one smooth pole and one singular one. This was the approach taken in \cite{Appels:2016uha, Appels:2017xoe, Gregory:2017ogk} where the North pole was takes to be regular, corresponding to the choice $K=1/P_-$ and clearly fixing $K$ in terms of the other parameters. In this work we will follow more closely in the footsteps of \cite{Anabalon:2018ydc, Anabalon:2018qfv, Cassani:2021dwa} where $K$ is allowed to remain generic and thus we allow for conical singularities at both poles.

\begin{figure}[ht] 
\begin{center}
\includegraphics[width=0.8\columnwidth]{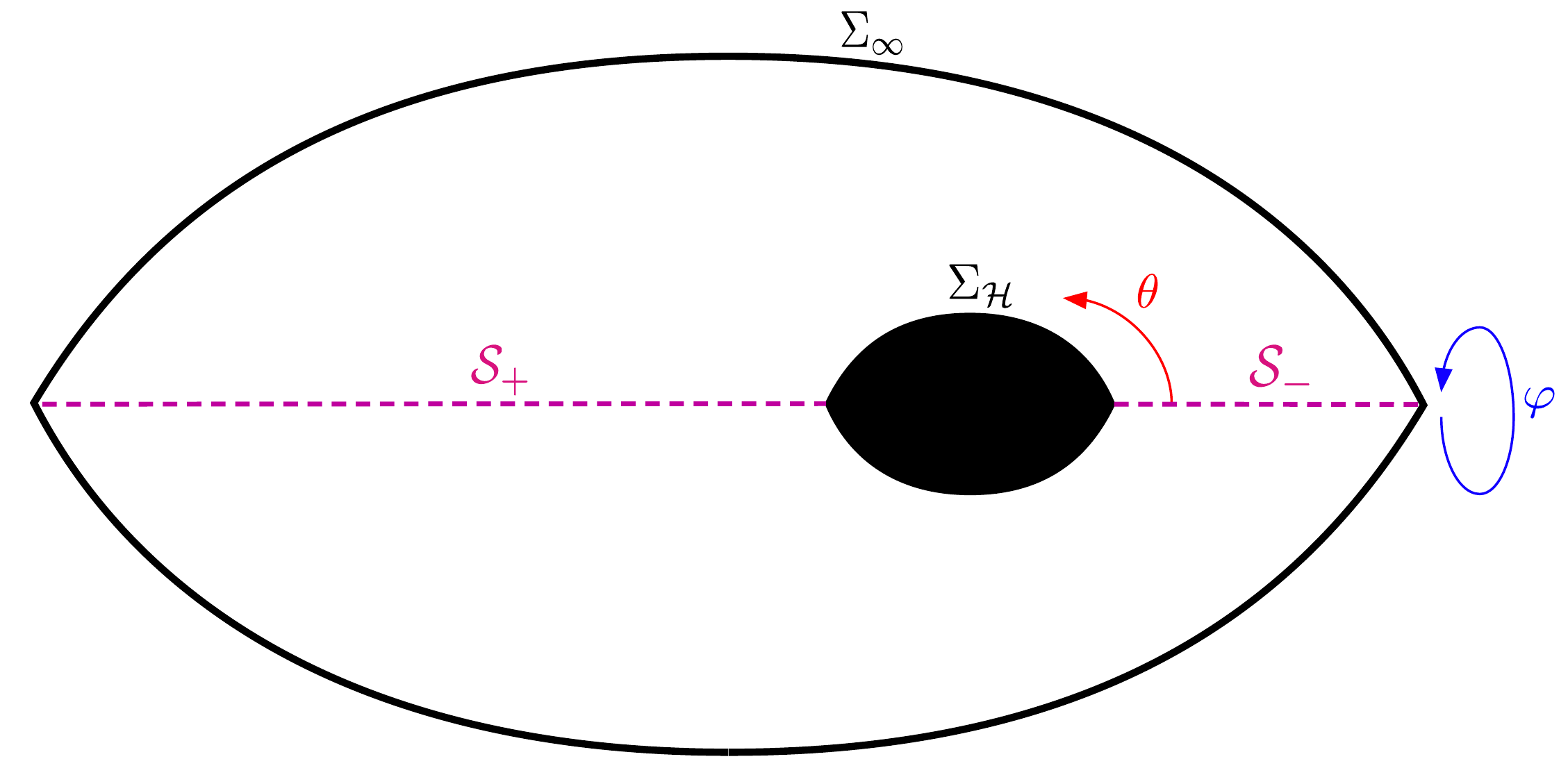}
\caption{A cartoon of a constant-$t$ slice of the accelerating black hole solution. The dark object in the interior is the black hole region with horizon cross-section at $r=r_+$ denoted by $\Sigma_{\mathcal{H}}$. Stretching from the horizon along the poles $\theta_{\pm}$ are two cosmic strings $\mathcal{S}_{\pm}$, providing conical deficits $\delta_{\pm}$ at the poles and physically understood to accelerate the black hole along the North pole axis, resulting in the black hole being moved from the ``centre" of the spacetime. The outer boundary $\Sigma_{\infty}$ is a cross-section of the conformal boundary $\mathscr{I}$. The axial coordinate $\varphi$ is suppressed in this picture which should be understood as a volume of revolution about the string axis.}
\label{fig: bh_cartoon}
\end{center}
\end{figure}

Physically, these singularities correspond to the presence of \textit{cosmic strings} stretching from the black hole horizon to conformal infinity, as shown in Fig. \ref{fig: bh_cartoon}. These cosmic strings have associated tensions given by
\begin{equation} \label{eq: cosmic_string_tensions}
    \mu_{\pm} = \frac{\delta_{\pm}}{8\pi G} = \frac{1}{4 G} (1 - P_{\pm}K),
\end{equation}
which \textit{accelerate} the black hole. We see explicitly that the overall tension is 
\begin{equation} \label{eq: overall_tension}
    \mu_- - \mu_+ = \frac{\alpha m K}{G} > 0 
\end{equation}
and thus the black hole accelerates in the North direction by virtue of $\alpha m K > 0$. We also note the value of the overall deficit in the spacetime 
\begin{equation} \label{eq: overall_deficit}
    \mu_- + \mu_+ = \frac{1}{2 G} (1- \Xi K),
\end{equation}
explicitly demonstrating that $K$ acts as a parameter for the conical deficit.

We finally note that we also require $P>0$ in order to have the correct signature of the full metric (\ref{eq: charged_acc_bh_metric}). As discussed in \cite{Anabalon:2018qfv, Ferrero:2020twa}, this means that we also have the following constraints between the parameters 
\begin{equation}
m \alpha < \begin{cases} \frac{\Xi}{2}  \quad \text{for} \quad \Xi \in (0,2], \\
 \sqrt{\Xi -1}  \quad \text{for} \quad \Xi >2,
\end{cases}
\end{equation}
which we will never use explicitly in any calculations in this paper, in a similar style to the assumption of slow acceleration.

\section{Asymptotic analysis} \label{sec: asymptotic_analysis}

In this section we will perform an asymptotic (i.e. near $\mathscr{I}$) analysis of the solution presented in equations (\ref{eq: charged_acc_bh_metric}) and (\ref{eq: charged_acc_bh_gauge_field}). This will allow us to demonstrate that the geometry is explicitly an asymptotically locally AdS (AlAdS) solution and, through the analysis of the variational problem, derive a constraint between the variations of the parameters. We also note that from this point on we will always use the normalisation of 
\begin{equation}
   \Lambda = -3 \iff \ell = 1,
\end{equation}
which can be reinstated via the usual dimensional analysis. 

For the asymptotic analysis of both the metric and the gauge field, we will often use the inverse radial coordinate $z>0$ of \cite{Cassani:2021dwa}, defined by 
\begin{equation} \label{eq: z_coord_def}
    \frac{1}{r} = \alpha \cos \theta + z,
\end{equation}
where $z = 0$ gives the location of $\mathscr{I}$ as this clearly corresponds to $H=0$.

\subsection{Gauge field}

The gauge field (\ref{eq: charged_acc_bh_gauge_field}) is smooth as one takes the limit $z = \epsilon \rightarrow 0$ and takes the boundary value 
\begin{equation} \label{eq: bdy_gauge_field}
 \mathbf{A}_{(0)} = \lim_{\epsilon \rightarrow 0} A_i|_{z = \epsilon} dx^i = - \cos \theta \left[ \frac{\alpha}{\kappa} e dt + g K d\varphi \right].
 \end{equation}
 This can be used to compute the boundary field strength via $\mathbf{F}_{(0)} = d\mathbf{A}_{(0)}$. In doing this, we note that we will sometimes switch between the usual azimuthal angle coordinate $\theta$ and an alternative coordinate $x$ given by
\begin{equation} \label{eq: azimuthal_coord_transform}
    x = \cos \theta
\end{equation}
and thus the boundary field strength takes the form
\begin{equation}
\mathbf{F}_{(0)} = \frac{\alpha e}{\kappa} dt \wedge dx - g K dx \wedge d\varphi. 
\end{equation}

The final asymptotic quantity that it will be important to define here is the electric current \cite{Cassani:2021dwa} 
\begin{equation}
    j^i  = - \frac{1}{4 \pi G} \lim_{\epsilon \rightarrow 0}  \left[ \frac{1}{\epsilon^3} n_{\mu} F^{\mu i} \right]_{z = \epsilon},
\end{equation}
where $n$ is the \textit{outward pointing} unit normal to the hypersurfaces of constant $z$. The only non-trivial components of the electric current are 
\begin{equation} \label{eq: current} 
    j^t = \kappa \frac{e}{4 \pi G}, \qquad j^{\varphi} = \frac{\alpha g}{4 K \pi G }.
\end{equation}

\subsection{Metric: Fefferman-Graham expansion}

We begin the asymptotic analysis of the metric (\ref{eq: charged_acc_bh_metric}) by providing the Fefferman-Graham expansion  \cite{Fefferman:1985zza}. This calculation has already been performed in \cite{Anabalon:2018ydc, Anabalon:2018qfv} (the asymptotic analysis was performed via an alternative  ADM approach in \cite{Cassani:2021dwa}) and here we will merely collect all of the prior results together and set our conventions. We begin by recalling that the Fefferman-Graham expansion for any AlAdS spacetime takes the form 
\begin{equation} \label{eq: FG_gauge}
ds^2 = \frac{1}{\rho^2} \left[ d\rho^2 + \left( g^{(0)}_{ij} + \rho^2 g^{(2)}_{ij} + \rho^3 g^{(3)}_{ij} + \ldots \right) dx^i dx^j \right],
\end{equation}
where $\rho > 0$ is an inverse radial coordinate and the conformal boundary $\mathscr{I}$ is located at $\rho =0$\footnote{Although $\mathscr{I}$ also corresponds $z=0$ as defined in (\ref{eq: z_coord_def}), $z \neq \rho$ away from $\mathscr{I}$. One can see this by applying the explicit coordinate transformation  (\ref{eq: z_coord_def}) to the metric (\ref{eq: charged_acc_bh_metric}).}. This gauge has proved extremely useful in studying AlAdS spacetimes in the AdS/CFT correspondence \cite{Henningson:1998gx, deHaro:2000vlm, Skenderis:2000in, Skenderis:2002wp}. The two key pieces of data in the expansion above are $g^{(0)}$ and $g^{(3)}$, which act as the CFT background metric and the expectation value of the CFT energy-momentum tensor respectively. The precise relationship \cite{deHaro:2000vlm} is 
\begin{equation}
     T_{ij}  =  \frac{3}{16 \pi G} g^{(3)}_{ij}.
\end{equation}

The explicit coordinate transformation which is required to put the metric (\ref{eq: charged_acc_bh_metric}) into the gauge (\ref{eq: FG_gauge}) was given in \cite{Anabalon:2018ydc, Anabalon:2018qfv} and for brevity we will not reproduce the steps here but merely summarise the important results of the expansion. Following the boundary coordinate convention of \cite{Cassani:2021dwa}, our chosen representative of the conformal class is given by 
 \begin{equation} \label{eq: bdy_metric_charged} 
    ds^2_{(0)} = - \frac{\tilde{P}}{\kappa^2} dt^2 + \frac{1}{P\tilde{P}} d\theta^2 + P K^2 \sin^2 \theta d\varphi^2, \qquad \tilde{P}(\theta) = 1 - \alpha^2 P(\theta) \sin^2 \theta.   \end{equation}
and the non-zero components of the energy-momentum tensor are (now using the coordinate $x$ defined in (\ref{eq: azimuthal_coord_transform})):
\begin{subequations}  \label{eq: T_charged} 
\begin{align}
   T^t_t  & = \frac{\left\{\alpha  m-2 (\Xi -1) x\right\} \left\{3 \alpha ^2 \left[x^2-1\right] \left[x (2 \alpha  m-\Xi  x+x)-1\right]-2\right\}}{8 \pi G \alpha }, \\
   T^x_x & = \frac{\alpha  m-2 (\Xi  -1)x}{8 \pi G  \alpha }, \\
   T^{\varphi}_{\varphi} & = -\frac{\left\{\alpha  m-2 (\Xi -1) x\right\} \left\{3 \alpha ^2 \left[x^2-1\right] \left[x (2 \alpha  m-\Xi  x+x)-1\right]-1\right\}}{8 \pi G \alpha }.
\end{align}
\end{subequations}
This formula is an extension of \cite{Anabalon:2018qfv} which now includes the magnetic charge parameter $g$. It can be obtained via the simple exchange of $e^2 \rightarrow e^2 +g^2$ in equation (A8) of that work. 

With all of the important boundary quantities defined, we note that a number of Ward identities are satisfied due to the bulk equations of motion. These take the form of conservation identities related to boundary diffeomorphisms and $U(1)$ gauge transformations respectively:
\begin{align} 
    \nabla^{(0)}_{i} T^i_j & = - j^i F^{(0)}_{ij} =  \frac{1-\Xi}{4\pi \alpha G} \delta^x_j, \label{eq: Conservation_WI} \\
    \nabla^{(0)}_i j^ i & = 0  \label{eq: Tracelessness_WI}.
\end{align}
where $\nabla^{(0)}$ is the Levi-Civita connection associated with (\ref{eq: bdy_metric_charged}) and all indices are understood to be moved with $g_{(0)}$. There is also a trace identity 
\begin{equation} \label{eq: T_tracelessness}
    T^i_i = \mathscr{A} = 0,
\end{equation}
where the right hand side of the above equation vanishes due to the vanishing of the trace anomaly $\mathscr{A}$ in four bulk dimensions \cite{Henningson:1998gx}.

\subsection{Boundary Cotton tensor} 

The boundary conformal class $[g_{(0)}]$ determines (in part) the asymptotic classification of the spacetime. In particular, we will follow \cite{Skenderis:2002wp, Papadimitriou:2004ap} in classifying a spacetime as \textit{asymptotically AdS} if $g_{(0)}$ is conformally flat and $\mathscr{I} \cong \mathbb{R} \times S^2$. If either of these conditions fail to hold then the spacetime will be \textit{asymptotically locally AdS} (AlAdS).

Restricting consideration to the case of $m \alpha \neq 0$\footnote{The $\alpha =0$ solutions are asymptotically AdS.}, the family of solutions (\ref{eq: charged_acc_bh_metric}) are AlAdS as they fail both of the criteria listed above. Firstly, we note that the presence of cosmic strings stretching from $\mathcal{H}$ to $\mathscr{I}$ gives a boundary topology of $\mathscr{I} \cong \mathbb{R} \times \Sigma_{\infty}$, where $\Sigma_{\infty}$ is a surface with one or two conical deficits due to the strings piercing the poles and thus the topological  condition is not satisfied. We will see that this plays a role in the construction of the conserved charges of these solutions in Section \ref{sec: Charges}. 

More importantly for our current analysis is the failure of the boundary to be conformally flat.
The conformal invariant we will use is the Cotton tensor of $g_{(0)}$, defined as 
\begin{equation} \label{eq: Cotton_def}
    C_{(0)}^{ij} = \varepsilon_{(0)}^{ikl} \nabla^{(0)}_k \left( R^{(0)j}_{\phantom{(0)}l} - \frac{1}{4} \delta^j_l R^{(0)} \right),
\end{equation}
where $\bm{\varepsilon}_{(0)}$ is the volume form for $g_{(0)}$, oriented as $\varepsilon^{(0)}_{t \theta \varphi} = \sqrt{-g_{(0)}}$. The Cotton tensor is symmetric and vanishes for any conformally flat 3-metric. Using the representative (\ref{eq: bdy_metric_charged}), we explicitly compute 
\begin{equation} \label{eq: Cotton_raised}
    C_{(0)}^{t\varphi} = C_{(0)}^{\varphi t} = \frac{6 \kappa  (\Xi-1) x -3 \alpha  \kappa  m}{K},
\end{equation}
which in particular is non-zero, demonstrating that the conformal boundary is not conformally flat and thus providing another criterion for this spacetime to be AlAdS. We conclude this subsection by noting that the tensor density $\sqrt{-g_{(0)}} C^{\phantom{(0)}i}_{(0) \phantom{i} j}$ is invariant under local conformal transformations, the non-trivial components of which are:
\begin{align}
    \sqrt{-g_{(0)}} C^{\phantom{(0)}t}_{(0) \phantom{t} \varphi} & =  3 K^2 \left(1-x^2\right)^{3/2} \left[\alpha  m-2 (\Xi -1) x\right] \left[x (2 \alpha  m-\Xi  x+x)-1\right], \label{eq: cotton_t_phi} \\
   \begin{split}
 \sqrt{-g_{(0)}} C^{\phantom{(0)}\varphi}_{(0) \phantom{\varphi} t} & = \frac{3 \sqrt{1-x^2} \left\{\alpha ^2 \left[x^2-1\right] \left[-2 \alpha  m x+(\Xi -1) x^2+1\right]+1\right\} }{\kappa ^2} \\ & \phantom{==} \times (2 ( 1- \Xi) x+\alpha  m). \label{eq: cotton_phi_t}
   \end{split}
\end{align}

It will be of interest to examine the allowed variations among the parameters $m, e, g, \alpha, K, \kappa$ which preserve the conformal class. Any variations which preserve the conformal class will be solutions to the equations 
\begin{equation} \label{eq: cotton_preservation}
    \delta \left ( \sqrt{-g_{(0)}} C^{\phantom{(0)}i}_{(0) \phantom{i} j} \right) = 0.
\end{equation}
In order to examine these solutions we start with the first non-trivial component (\ref{eq: cotton_t_phi}) and note that (\ref{eq: cotton_preservation}) has to be satisfied at all orders in $x$. Ignoring the factor of $(1-x^2)^{3/2}$ which sits out of the front of the equation as an overall phase space constant, we find the following equations at each order in $x$
\begin{align}
    \mathcal{O}(x^0) : \quad  0 & =  \delta ( K^2 m \alpha ), \\
   \mathcal{O}(x^1) : \quad 0 & = \delta [K^2 (\Xi -1 + m^2 \alpha^2) ], \\
   \mathcal{O}(x^2) : \quad 0 & = \delta [K^2 m \alpha (\Xi -1) ], \\
   \mathcal{O}(x^3) : \quad 0 & = \delta [ K (\Xi -1) ], 
\end{align}
which can be solved simultaneously to give 
\begin{equation} \label{eq: first_cotton_relations} 
    \delta K = \delta \Xi = \delta (m \alpha ) = 0. 
\end{equation}
We now look at the variation of the other non-trivial component, i.e. the one given in equation (\ref{eq: cotton_phi_t}). This equation will produce equations at five different orders in $x$ (ignoring the overall $\sqrt{1-x^2}$ factor) but we will only need two of them, namely 
\begin{align}
     \mathcal{O}(x^0): \quad 0 & = \delta \left( \frac{m \alpha (-1 + \alpha^2 )}{\kappa^2} \right) \implies \delta \left( \frac{-1 + \alpha^2}{\kappa^2} \right) = 0, \\
     \mathcal{O}(x^4) : \quad 0 & = \delta  \left( \frac{m \alpha^3 (\Xi - 1)}{\kappa^2} \right) \implies \delta \left( \frac{\alpha^2}{\kappa^2} \right) = 0, 
\end{align}
where we used the relations in (\ref{eq: first_cotton_relations}). Subtracting these two equations leaves us with the result for the allowed variations of the parameters
\begin{equation} \label{eq: cotton_pres_constraints}
    \delta \kappa = \delta (m \alpha  ) = \delta \alpha = \delta K = \delta \Xi = 0. 
\end{equation}

Despite this seeming like a very strong restriction upon the space of parameters, we note that $\delta \Xi = 0$ does not entirely fix the electric and magnetic parameters $e$ and $g$. Instead it allows for a circle on the phase space 
\begin{equation} \label{eq: em_circle}
    e^2 + g^2 = c^2, 
\end{equation}
where $c$ is a phase space constant. This is to be expected as $e, g$ only enter the metric via $\Xi$, so analysis of the metric will not put any constraints upon them individually. We will return to analyse the variations of the gauge parameters in the next section.

\subsection{Variational problem} \label{sec: variational_principle}

We analyse the variational problem for the family of spacetimes (\ref{eq: charged_acc_bh_metric}), an issue which will be crucial in determining the class of variations which are allowed to enter into the first law of accelerating black hole thermodynamics. We begin by noting that the bulk action (\ref{eq: bulk_action}) must first be supplemented by a boundary action consisting of the Gibbons-Hawking-York boundary term as well as the usual holographic counterterms \cite{Balasubramanian:1999re, deHaro:2000vlm, Skenderis:2000in, Skenderis:2002wp}. We begin by presenting this action at a regulated boundary $z=\epsilon >0$ 
\begin{equation} \label{eq: bdy_counterterms}
    S_{\text{bdy}} = S_{\text{GHY}} + S_{\text{ct}} = \frac{1}{16\pi G } \int_{z =\epsilon} d^3 x \, \sqrt{-h} (2 \mathcal{K} - 4 - \mathcal{R} ),
\end{equation}
where $h_{ij}$ is the induced metric on the hypersurface $z=\epsilon$, $\mathcal{K}$ is its trace extrinsic curvature when embedded in the bulk spacetime and $\mathcal{R}$ is the Ricci scalar of $h$. The total action is thus\footnote{One may conjecture that the presence of the cosmic strings should also alter the action. However, this has already been shown not to be the case \cite{Gregory:2001dn} as the delta function arising from the extrinsic curvature precisely cancels the string contribution.}
\begin{equation} \label{eq: S_reg}
S = S_{\text{bulk}} + S_{\text{bdy}} 
\end{equation}
and we define the renormalised action as 
\begin{equation}
S_{\text{ren}} =  \lim_{\epsilon \rightarrow 0} S.
\end{equation}

The variational problem is well-posed when variations of the renormalised action vanish iff the equations of motion are satisfied. The general formula for the variation of the renormalised on-shell action is \cite{Papadimitriou:2005ii}
\begin{equation} \label{eq: on_shell_variation}
    \delta S_{\text{ren}}  \approx \int_{\mathscr{I}} d^3 x \sqrt{-g_{(0)}} \, \left( \frac{1}{2} T^{ij} \delta g^{(0)}_{ij} + j^i \delta A^{(0)}_i  \right)
\end{equation}
and thus the variational problem is well-posed when the right hand term above vanishes. The typical way of ensuring this is to select Dirichlet boundary conditions \cite{Papadimitriou:2005ii}, i.e. to demand that the variations satisfy 
\begin{equation} \label{eq: Dirichlet_bc}
    \delta g^{(0)}_{ij} \propto g^{(0)}_{ij}, \qquad \delta A^{(0)}_i = 0,
\end{equation}
which clearly makes the variational problem well-posed due to the tracelessness condition (\ref{eq: T_tracelessness}). We will now show explicitly that these boundary conditions cannot be satisfied non-trivially for the solutions (\ref{eq: charged_acc_bh_metric}) when we treat the parameters $m, e, g, \alpha, K, \kappa$ as changing under variation. 

We start with the metric boundary condition in (\ref{eq: Dirichlet_bc}) which we have already built up to in the analysis of the Cotton tensor (\ref{eq: Cotton_raised}) in the previous subsection. Due to the fact that $\sqrt{-g_{(0)}} C^{\phantom{(0)}i}_{(0) \phantom{i} j}$ is invariant under local conformal transformations and varies under changes in the conformal class, we have the following relationship amongst the variations:
\begin{equation}
     \delta g^{(0)}_{ij} \propto g^{(0)}_{ij} \iff   \delta \left ( \sqrt{-g_{(0)}} C^{\phantom{(0)}i}_{(0) \phantom{i} j} \right)  = 0 ,
\end{equation}
where we have already determined that space of variations in the right hand set above are those given in equations (\ref{eq: cotton_pres_constraints}) and (\ref{eq: em_circle}). 

Moving on to the gauge field, we start by using (\ref{eq: bdy_gauge_field}) to write the constraint of \cite{Papadimitriou:2005ii} as 
\begin{equation} \label{eq: bc_gauge_field}
  0 = \delta \mathbf{A}_{(0)} = \delta \left( - \cos \theta \left[ \frac{\alpha}{\kappa} e dt + g K d\varphi \right] \right),
\end{equation}
resulting in
\begin{equation}
    \delta e = \delta g = 0,
\end{equation}
which we note is a much stronger constraint than the one imposed for the solutions with $\alpha = g = 0$ (such as those considered in \cite{Papadimitriou:2005ii}), where the boundary condition (\ref{eq: bc_gauge_field}) is satisfied trivially.
This analysis demonstrates that if one wants to apply Dirichlet boundary conditions (\ref{eq: Dirichlet_bc}) then the only allowed perturbation is the trivial one 
\begin{equation}
    \delta m = \delta \alpha = \delta K = \delta e = \delta g = \delta \kappa = 0. 
\end{equation}

At first sight this appears to be a troubling result. It means that studying the black hole first law for the class of accelerating solutions (\ref{eq: charged_acc_bh_metric}) via the approach of \cite{Papadimitriou:2005ii} is not possible. Indeed, this tension was remarked upon in \cite{Cassani:2021dwa} due to the variations in that work changing the boundary conformal class $[g_{(0)}]$. It is clear that if we wish to apply techniques along the line of \cite{Papadimitriou:2005ii} then we need to consider more general boundary conditions than the Dirichlet conditions (\ref{eq: Dirichlet_bc}). 

In order to resolve these tensions, one may take inspiration from \cite{Anabalon:2018qfv}, where the requirement of a well-posed variational problem was used to fix $\kappa$ in terms of the other parameters. We follow \cite{Compere:2008us} in considering the most general solutions to the variational problem i.e. we look to solve
\begin{equation} \label{eq: variational_principle_well_posed}
     \delta S_{\text{ren}}  \approx \int_{\mathscr{I}} d^3 x \sqrt{-g_{(0)}} \, \left( \frac{1}{2} T^{ij} \delta g^{(0)}_{ij} + j^i \delta A^{(0)}_i  \right) = 0,
\end{equation}
without applying any specific boundary conditions upon the metric and gauge field. The analysis of the $U(1)$ term goes through very straightforwardly due to the gauge choice (\ref{eq: charged_acc_bh_gauge_field}): applying equations (\ref{eq: bdy_gauge_field}) and (\ref{eq: current}), we obtain 
\begin{equation}
    \sqrt{-g_{(0)}} j^i \delta A^{(0)}_i = - \frac{\sin 2 \theta}{8 \pi G} \frac{K}{\kappa} \left[ e \kappa \delta \left( \frac{\alpha e}{\kappa} \right) + g \alpha \delta (g K) \right],
\end{equation}
which vanishes under the integration performed in (\ref{eq: variational_principle_well_posed}), reducing that equation to a purely gravitational problem, i.e. we only need to solve
\begin{equation}
     \delta S_{\text{ren}}  \approx \frac{1}{2} \int_{\mathscr{I}} d^3 x \sqrt{-g_{(0)}} \,   T^{ij} \delta g^{(0)}_{ij} = 0.
\end{equation}
Using (\ref{eq: bdy_metric_charged}) and (\ref{eq: T_charged}) and performing the $\theta$ (equivalently $x$) integral we reach
\begin{equation} \label{eq: variational_constraint_general}
     - 2 K \alpha \kappa \Xi \delta \alpha + 2 K (\alpha^2 \Xi -1) \delta \kappa + \kappa( 2\alpha^2 \Xi -1) \delta K = 0,
\end{equation}
which is the master formula constraining the variations of the parameters in order to have a well-posed variational problem. We will use this relation in order to derive the first law. 

\subsection{Comment on the ``master formula''}

We conclude this section with some comments concerning our master formula (\ref{eq: variational_constraint_general}) and a comparison to previous literature \cite{Anabalon:2018qfv, Cassani:2021dwa, Barrientos:2022bzm}. The first point we note about (\ref{eq: variational_constraint_general}) is that it is not an integrable equation on the phase space i.e. one cannot directly solve for $\kappa$ as a function of the other parameters. In order to examine the nature of the non-integrability, we find it useful to first manipulate equation (\ref{eq: variational_constraint_general}) into the following form
\begin{equation} \label{eq: variational_constraint_manipulated}
    \delta \log \kappa - \frac{1}{2} \delta \log \left( \Xi (1 - \alpha^2  \Xi ) \right) = \frac{  1- 2\alpha^2 \Xi}{2 \alpha^2 \Xi -2 }  \delta \log(\Xi K),
\end{equation}
where we note that the terms on the left hand side are exact 1-forms on the phase space and thus the integrability is spoiled entirely by the term on the right hand side. In order to recover integrability, we immediately see that one possibility is to consider a restricted set of variations which satisfy
\begin{equation} \label{eq: kappa_old_constraint}
    \delta ( \Xi K) = 0,
\end{equation}
a constraint which can be interpreted as fixing the overall deficit in the spacetime (\ref{eq: overall_deficit}). Upon making this choice, one can then solve (\ref{eq: variational_constraint_manipulated}) for $\kappa$ directly in order to find 
\begin{equation} \label{eq: kappa_old}
    \kappa = \sqrt{\Xi(1-\alpha^2\Xi)},
\end{equation}
is agreement with the $\kappa$ found in previous work \cite{Appels:2016uha, Appels:2017xoe, Gregory:2017ogk, Anabalon:2018ydc, Anabalon:2018qfv, Cassani:2021dwa}. This also agrees with the motivation of \cite{Anabalon:2018qfv, Cassani:2021dwa, Barrientos:2022bzm}, where the requirement of a well-posed variational problem (in the setting of fixed cosmic string tensions) was used to derive $\kappa$. It is interesting to note that (\ref{eq: kappa_old_constraint}) is actually a smaller set of constraints than those postulated in \cite{Anabalon:2018qfv, Cassani:2021dwa, Barrientos:2022bzm}, where both tensions $\mu_{\pm}$ were chosen to be fixed, rather than just their sum.

Returning to equation (\ref{eq: variational_constraint_manipulated}), we note that although the non-integrability may appear at first to be a disturbing feature, there is no a priori reason for (\ref{eq: variational_constraint_manipulated}) to be an integrable equation as the integrand in (\ref{eq: variational_principle_well_posed}) is not a closed form on the phase space. Schematically speaking, we start with a generic six parameter phase space $\{m, e, g, \alpha, K, \kappa\}$ and the imposition of well-posedness generically leads to a differential equation containing variations of a subset of these parameters. This has the effect of reducing the number of degrees of freedom to five, without providing an explicit relation due to the non-integrability. 

In order to recover integrability one must necessarily consider a restricted set of variations such as those given in equation (\ref{eq: kappa_old_constraint}), resulting in a restricted phase space. Working more generally, we note that the non-integrable master formula (\ref{eq: variational_constraint_manipulated}) illustrates that $\kappa$ as given in (\ref{eq: kappa_old}) is the correct choice when one fixes the overall deficit. However, if one chooses $\kappa$ as given in (\ref{eq: kappa_old}) but does not impose (\ref{eq: kappa_old_constraint}) then the variational problem is generically \textit{ill-posed}. The general philosophy (and technical tool) is simply that one needs to consider a space of variations amongst the parameters for which the variational problem (\ref{eq: variational_principle_well_posed}) is well-posed. This new approach to the accelerating solutions allows us to study thermodynamics for (many) alternative choices of $\kappa$. In proceeding, we will work completely generically, i.e. we will not fix $\kappa$ to the form of (\ref{eq: kappa_old}) but rather we will merely require that the master constraint equation (\ref{eq: variational_constraint_general}) holds. We will now construct the conserved charges and derive the most general form of the first law for which the variations are well-posed and the cosmic string tensions $\mu_{\pm}$ are independent thermodynamic variables. In doing this, we will regularly highlight the importance of well-posedness.

\section{Charges} \label{sec: Charges}

Before discussing the thermodynamics of the solution specified by (\ref{eq: charged_acc_bh_metric}) and (\ref{eq: charged_acc_bh_gauge_field}), we first need to establish the appropriate conserved charges which will later appear in the first law. In order to define these charges we utilise the \textit{covariant phase space} formalism, following closely in the style of \cite{Crnkovic:1986ex, Lee:1990nz, Wald:1993nt, Iyer:1994ys, Iyer:1995kg, Wald:1999vt, Wald:1999wa}. We will now briefly review this formalism for a generic diffeomorphism covariant Lagrangian theory in $d$-dimensions in order to familiarise the reader with the notation, before applying the tools to our theory of interest (\ref{eq: bulk_action}).

\subsection{Covariant phase space}

We begin by considering a variation of the Lagrangian $d$-form $\mathbf{L}[\psi]$, where $\psi$ denotes the dynamical fields of the theory. A variation of $\mathbf{L}$ takes the generic form 
\begin{equation}
    \delta \mathbf{L} = \mathbf{E}[\psi] \delta \psi + d \bm{\Theta}[\psi; \delta \psi],
\end{equation}
where $\mathbf{E}$ is the equation of motion $d$-form ($\mathbf{E} \approx 0$) and $\bm{\Theta}$ is a $(d-1)$-spacetime form called the \textit{symplectic potential}.\footnote{Strictly speaking, this object is a \textit{pre}-symplectic potential as there is no guarantee that it will give rise to an invertible symplectic form on phase space. We will ignore these subtleties as they will play no role in our analysis.} We note that $\bm{\Theta}$ is also a 1-form on the phase space. 

Using $\bm{\Theta}$, one can construct the \textit{symplectic current}
\begin{equation} \label{eq: omega}
    \bm{\omega}[\psi; \delta_1 \psi, \delta_2 \psi] = \delta_{1} \bm{\Theta}[\psi; \delta_{2} \psi] - \delta_{2} \bm{\Theta}[\psi; \delta_{1} \psi], 
    \end{equation}
a $(d-1)$-form on spacetime and a 2-form on phase space. Integrating the symplectic current over a partial Cauchy slice $C$ defines the \textit{symplectic form} 
\begin{equation} \label{eq: symplectic_form}
\Omega_C(\psi; \delta_1 \psi, \delta_2 \psi) = \int_{C} \bm{\omega}[\psi; \delta_1 \psi, \delta_2 \psi],
\end{equation}
a spacetime scalar and a phase space 2-form. 

In order to construct the charges, we will often be interested in the case when one of the variations is generated by a Killing vector field $\xi$ ($\delta_{\xi} = \mathcal{L}_{\xi}$) or a $U(1)$ gauge transformation $f$. For these cases we are able to define the \textit{Wald Hamiltonians} corresponding to these transformations via 
\begin{equation} \label{eq: Wald_Hamiltonian_diffeo}
    \delta H_{\xi} = \Omega_C(\psi; \delta \psi, \mathcal{L}_{\xi} \psi) = \int_{C} \bm{\omega}[\psi; \delta \psi, \mathcal{L}_{\xi} \psi]
\end{equation}
and 
\begin{equation} \label{eq: Wald_Hamiltonian_gauge}
    \delta H_{f} = \Omega_C(\psi; \delta \psi, \delta_{f} \psi) = \int_{C} \bm{\omega}[\psi; \delta \psi, \delta_{f} \psi].
\end{equation}

It remains to be seen that these Wald Hamiltonians are boundary quantities. We show this first for $H_{\xi}$ by defining the Noether current $(d-1)$-form 
\begin{equation}
\mathbf{J}[\xi] = \bm{\Theta}[\psi; \mathcal{L}_{\xi} \psi] - i_{\xi} \mathbf{L},
\end{equation}
an object which is locally exact on-shell i.e. when $\mathbf{E} = 0$, we have 
\begin{equation}
\mathbf{J}[\xi] = d \mathbf{Q}[\xi],
\end{equation}
where $\mathbf{Q}$ is the \textit{Noether charge} $(d-2)$-form. Restricting to the case when both the equations of motion $\mathbf{E}=0$ and the linearised equations of motion $\delta \mathbf{E} = 0$ are satisfied, one is able to show that \cite{Iyer:1994ys, Papadimitriou:2005ii} 
\begin{equation} \label{eq: omega_exact}
\bm{\omega}[\psi; \delta \psi, \mathcal{L}_\xi \psi] = d \left( \delta  \mathbf{Q}[\xi] - i_{\xi} \bm{\Theta}[\psi; \delta \psi] \right)
\end{equation}
and thus 
\begin{equation} \label{eq: diff_Hamiltonian_divergent}
 \delta H_{\xi} = \Omega_C(\psi; \delta \psi, \mathcal{L}_{\xi} \psi) = \int_{C} \bm{\omega}[\psi; \delta \psi, \mathcal{L}_{\xi} \psi] = \int_{\partial C_{\infty}} \delta  \mathbf{Q}[\xi] - i_{\xi} \bm{\Theta}[\psi; \delta \psi],
\end{equation}
where $\partial C_{\infty}$ is the intersection of $C$ with the conformal boundary.

For the $U(1)$ transformation, the analysis is even simpler in that the Noether current takes the form 
\begin{equation}
\mathbf{J}[f] = \bm{\Theta}[\psi; \delta_{f} \psi] \approx d \mathbf{Q}[f].
\end{equation}
By gauge invariance of the symplectic potential we have
\begin{equation}
\bm{\omega}[\psi; \delta \psi, \delta_{f} \psi] = \delta \bm{\Theta}[\psi; \delta_{f} \psi] \approx d \delta \mathbf{Q}[f],
\end{equation}
and thus the Wald Hamiltonian is 
\begin{equation}
\delta H_{f} = \Omega_C(\psi; \delta \psi, \delta_{f} \psi) = \int_{C} \bm{\omega}[\psi; \delta \psi, \delta_{f} \psi] = \int_{\partial C_{\infty}} \delta  \mathbf{Q}[f]. 
\end{equation}
This can be immediately integrated on phase space to give
\begin{equation} \label{eq: electric_Hamiltonian}
    H_{f} = \int_{\partial C_{\infty}} \mathbf{Q}[f].
\end{equation}
\subsection{Corner improvement}

With the general terminology of the covariant phase space now introduced, we now study Einstein-Maxwell theory (\ref{eq: bulk_action}) using these techniques. We return to working explicitly in $d=4$ and note that the dynamical fields of this theory amount to 
\begin{equation}
\psi = \{ g_{\mu \nu}, A_{\rho}  \}. 
\end{equation}
 The variations of the metric under the diffeomorphism and $U(1)$ transformations are
\begin{equation} \label{eq: metric_transformation}
\delta_{\xi} g_{\mu \nu} = \mathcal{L}_{\xi} g_{\mu \nu}, \qquad \delta_{f} g_{\mu \nu} = 0,  
\end{equation}
and the variations of the gauge field are
\begin{equation}
\delta_{\xi} A_{\mu} = \mathcal{L}_{\xi} A_{\mu} , \qquad \delta_{f} A_{\mu} = \partial_{\mu} f,
\end{equation}
where we note that we have taken the diffeomorphism to also act on the gauge field with the Lie derivative. This is in order to preserve our gauge choice (\ref{eq: charged_acc_bh_gauge_field}), which will turn out to be a particularly convenient choice to analyse the charges and thermodynamics of the solution. One can instead work in an entirely gauge-independent manner \cite{Elgood:2020svt, Elgood:2020mdx, Elgood:2020nls, Ortin:2022uxa} in which case the diffeomorphism transformation above includes an additional $\xi$-dependent gauge transformation: $\delta_{\xi} A_{\mu} =  \mathcal{L}_{\xi} A_{\mu} + d \chi_{\xi}$. When restricting to $\xi$ as a Killing vector, our gauge choice (\ref{eq: charged_acc_bh_gauge_field}) is such that $\mathcal{L}_{\xi} A_{\mu} = 0$ and thus the only allowed gauge transformations would be $\chi_{\xi} = a_1 dt + a_2 d\varphi$, with $a_{1,2}$ constants. These transformations have no effect on the laws of thermodynamics \cite{Cassani:2021dwa} and thus we will use the simpler transformation formulae above. 

We are almost at the stage of being able to compute the conserved quantities which will appear in the first law. There is one final subtlety for the accelerating solutions (\ref{eq: charged_acc_bh_metric}), (\ref{eq: charged_acc_bh_gauge_field}) in that the cross-sections of the conformal boundary $\partial C_{\infty} = \Sigma_{\infty}$ themselves have boundary: $\partial \Sigma_{\infty} = S^1_{-} \sqcup S^1_+$, where $S^1_{\pm}$ are the small circles around the cosmic strings at $\theta_{\pm}$ respectively. The inclusion of these boundaries will mean that equation (\ref{eq: diff_Hamiltonian_divergent}) needs to be supplemented by a \textit{corner improvement} in order to give the correct charges.

The form of this corner term can be discerned from the holographic counterterms (\ref{eq: bdy_counterterms}). The counterterms are not only responsible for renormalising the action but also the symplectic potential \cite{Papadimitriou:2005ii}. The full expression for the renormalised symplectic potential, including the corner terms, is given in equation (2.13) of \cite{Compere:2020lrt} and in our notation reads
\begin{equation} \label{eq: ren_theta}
\bm{\Theta}_{\text{ren}}[\psi; \delta \psi] \equiv \bm{\Theta}[\psi; \delta \psi] - \delta \mathbf{L}_{\text{GHY}}[h_{ij}] - \delta \mathbf{L}_{\text{ct}}[h_{ij}] + d \bm{\Theta}_{\text{ct}}[h_{ij}; \delta h_{ij}],
\end{equation}
where $\mathbf{L}_{\text{ct}}$ is the counterterm Lagrangian and $\bm{\Theta}_{\text{ct}}$ is the symplectic potential arising from the variation of counterterm action, defined via
\begin{equation}
    \delta \mathbf{L}_{\text{ct}}[h_{ij}] = \mathbf{E}_{(3)}^{kl}[h_{ij}] \delta h_{kl} + d \bm{\Theta}_{\text{ct}}[h_{ij}; \delta h_{ij}],
\end{equation}
where $\mathbf{E}_{(3)}^{ij}$ are the ``equations of motion" for the boundary metric $h_{ij}$, explicitly not satisfied as we do not impose dynamical gravity on the boundary. 

The adjustments to $\bm{\Theta}$ in (\ref{eq: ren_theta}) can be seen as utilising all of the inherent ambiguities in the construction of the symplectic potential \cite{Iyer:1994ys}. The $\delta \mathbf{L}_{\text{bdy}} = \delta \mathbf{L}_{\text{GHY}}+\delta \mathbf{L}_{\text{ct}}$ term does not alter the symplectic current (\ref{eq: omega}) and as such it will not affect the either the symplectic form (\ref{eq: symplectic_form}) or the Wald Hamiltonians (\ref{eq: Wald_Hamiltonian_diffeo}), (\ref{eq: Wald_Hamiltonian_gauge}). We can safely ignore the contribution from this term in our analysis. 

The $d \bm{\Theta}_{\text{ct}}$ term will be important. In order to see this, we recall that this $d$-exact term shifts the Noether charge form \cite{Iyer:1994ys} as
\begin{equation}
    \mathbf{Q}_{\text{ren}} = \mathbf{Q} + \bm{\Theta}_{\text{ct}}[h_{ij}; \mathcal{L}_{\xi} h_{ij}]
\end{equation}
and thus the renormalised diffeomorphism Hamiltonian is 
\begin{align}
\begin{split}
\delta H^{\text{ren}}_{\xi}  & = \delta H_{\xi} + \int_{\Sigma_{\infty}} \delta \bm{\Theta}_{\text{ct}}[h_{ij}; \mathcal{L}_{\xi} h_{ij}] - \mathcal{L}_{\xi} \bm{\Theta}_{\text{ct}}[h_{ij}; \delta h_{ij}] + d i_{\xi} \bm{\Theta}_{\text{ct}}[h_{ij}; \delta h_{ij}] \\
& = \delta H_{\xi} + \int_{\Sigma_{\infty}} \bm{\omega}_{\text{ct}}[h_{ij}; \delta h_{ij}, \mathcal{L}_{\xi} h_{ij}]  + d i_{\xi} \bm{\Theta}_{\text{ct}}[h_{ij}; \delta h_{ij}],
\end{split}
\end{align}
where we used the form expression for the Lie derivative $\mathcal{L}_{\xi} = i_{\xi} d + d i_{\xi}$ and introduced the counterterm symplectic current $\bm{\omega}_{\text{ct}}$ in the second line. If we restrict consideration to the case of $\xi$ being an asymptotic Killing vector \cite{Papadimitriou:2005ii}, then the vector will preserve the conformal class $[g_{(0)}]$ and thus 
\begin{equation}
    \left. \bm{\omega}_{\text{ct}}[h_{ij}; \delta h_{ij}, \mathcal{L}_{\xi} h_{ij}] \right|_{\Sigma_{\infty}} = 0,
\end{equation}
which allows us to write our final formula for the renormalised Hamiltonian as 
\begin{equation} \label{eq: H_ren}
    \delta H^{\text{ren}}_{\xi} =  \delta H_{\xi} + \int_{\partial \Sigma_{\infty}}  i_{\xi} \bm{\Theta}_{\text{ct}}[h_{ij}; \delta h_{ij}],
\end{equation}
a formula which can be viewed as the extension of those in \cite{Papadimitriou:2005ii, Compere:2020lrt} to encompass spacetimes where cross-sections of $\mathscr{I}$ have non-vanishing boundary. The new term acts as a counterterm at $\mathcal{O}(1/\rho)$ in the coordinates (\ref{eq: FG_gauge}) and plays a similar role to the counterterm required to define charges in NUT charged spacetimes \cite{Godazgar:2022jxm}. This is perhaps expected as in that case the presence of \textit{Misner strings} results in singularities in much the same way the cosmic strings do in our setup. We will see further similarities between these cases when discussing the first law of thermodynamics.

We conclude this section by noting that due to the transformation properties of the metric (\ref{eq: metric_transformation}) we have $\delta_{f} h_{ij} = 0$ and thus the $U(1)$ Hamiltonian is invariant i.e.
\begin{equation} \label{eq: H_alpha_ren}
     H_{f}^{\text{ren}} = H_{f}.
\end{equation}
We will now apply formulae (\ref{eq: H_ren}) and (\ref{eq: H_alpha_ren}) to compute the charges for our accelerating solution (\ref{eq: charged_acc_bh_metric})-(\ref{eq: charged_acc_bh_gauge_field}).

\subsection{Mass}

The first charge we compute is the mass charge $\mathcal{M}$, given by 
\begin{equation}
\mathcal{M} = H^{\text{ren}}_{\xi},
\end{equation}
where we take the timelike Killing vector to be
\begin{equation} \label{eq: xi_killing}
 \xi =  \partial_t.
\end{equation}
Note that this is the same normalisation as \cite{Anabalon:2018ydc, Anabalon:2018qfv, Cassani:2021dwa}, where they argued that the normalisation was crucial in arriving at the correct law of thermodynamics. The key difference in our approaches is that we use the well-posedness master equation (\ref{eq: variational_constraint_general}) as our guiding principle, and thus we will arrive at the correct first law for \textit{any parameter independent} normalisation, as long as equation (\ref{eq: variational_constraint_general}) is satisfied. If the normalisation depends upon the parameters we no longer have $\delta \xi = 0$, which is an important assumption in \cite{Wald:1993nt}. Allowing for ``field dependent"  symmetries in the formalism is a topic of some study, (see e.g. related formulae in \cite{Compere:2019bua, Ball:2020vzo}) but we do not consider such cases here.

We will be explicit in constructing the mass charge as it will provide an important illustration of the corner improvement present in our formula (\ref{eq: H_ren}). We begin by noting that one could completely bypass this discussion of covariant phase space/Wald Hamiltonians and simply use the \textit{holographic mass} formulae of \cite{Hollands:2005wt, Papadimitriou:2005ii}
\begin{equation} \label{eq: holographic_mass}
    \mathcal{M}_{\text{hol}} = - \int_{\Sigma_{\infty}} d^2 x \, \sqrt{-g_{(0)}} \left( T^t_i + j^t A^{\text{(0)}}_i \right) \xi^i,
\end{equation}
which was the approach taken in \cite{Anabalon:2018ydc, Anabalon:2018qfv, Cassani:2021dwa}. Putting our vector (\ref{eq: xi_killing}) into the above formula and using the gauge choice (\ref{eq: bdy_gauge_field}) we find the holographic mass to be
\begin{equation}
 \mathcal{M}_{\text{hol}} =  - \int_{\Sigma_{\infty}} d^2 x  \, \sqrt{-g_{(0)}} T^t_t = \frac{Km(1-\alpha^2 \Xi)}{G \kappa}.
\end{equation}

It was shown in \cite{Papadimitriou:2005ii} that the holographic mass was equivalent to the Wald Hamiltonian $H_{\xi}$ for spacetimes without conical deficit. Here we extend this proof to include those which do. We start by computing the contribution to (\ref{eq: H_ren}) from the first term, i.e. the ``bare" Wald Hamiltonian 
\begin{equation} \label{eq: bare_Hamiltonian}
    \delta H_{\xi} =  \int_{\Sigma_{\infty}} \delta  \mathbf{Q}[\xi] - i_{\xi} \bm{\Theta}[\psi; \delta \psi],
\end{equation}
where 
\begin{equation} \label{eq: symp_split}
\mathbf{Q} = \mathbf{Q}_{\text{EH}} + \mathbf{Q}_{\text{M}}, \qquad \mathbf{\Theta} = \mathbf{\Theta}_{\text{EH}} + \mathbf{\Theta}_{\text{M}}
\end{equation}
for the respective Einstein-Hilbert and Maxwell contributions to the bulk action (\ref{eq: bulk_action}). The formulae for the symplectic quantities for these theories are well-known, see for example \cite{Iyer:1994ys, Papadimitriou:2005ii, Ball:2020vzo} and read as follows\footnote{The convention for the volume form is 
\begin{equation*} \label{eq: volume_form_r}
    \varepsilon_{tr\theta\varphi} = \sqrt{- g } = \frac{r^2 \sin \theta}{H^4 \kappa} K.
\end{equation*}} 
\begin{align}
      \mathbf{Q}_{\text{EH}} & = \frac{1}{16 \pi G} \cdot \frac{1}{2!} \varepsilon_{\mu \nu \rho \sigma}  \nabla^{\nu} \xi^{\mu} dx^{\rho} \wedge dx^{\sigma}, \label{eq: Q} \\
    \mathbf{\Theta}_{\text{EH}} & = \frac{1}{16 \pi G} \cdot \frac{1}{3!} \varepsilon_{\mu \nu \rho \sigma} \left( \nabla^{\mu} (g_{\alpha \beta} \delta g^{\alpha \beta} ) - \nabla_{\alpha} \delta g^{\mu \alpha} \right) dx^{\nu} \wedge dx^{\rho} \wedge dx^{\sigma}, \\
     \mathbf{Q}_{\text{M}} & = - \frac{1}{4 \pi G} (i_{\xi} \mathbf{A} ) * \mathbf{F} = - \frac{1}{4 \pi G} \cdot \frac{1}{2!} \xi^{\mu} A_{\mu}  (* \mathbf{F})_{\rho \sigma} dx^{\rho} \wedge dx^{\sigma}, \label{eq: Theta_M}  \\
    \mathbf{\Theta}_{\text{M}} & =  - \frac{1}{4\pi G} \delta \mathbf{A} \wedge * \mathbf{F} = - \frac{1}{4\pi G} \cdot \frac{1}{2!} \delta A_{\nu} (* \mathbf{F})_{\rho \sigma} dx^{\nu} \wedge dx^{\rho} \wedge dx^{\sigma},
\end{align}
which agree with the formulae of \cite{Ball:2020vzo} up to our differing normalisations of the gauge fields. We can quickly see that all of the Maxwell terms in the charge integral (\ref{eq: bare_Hamiltonian}) drop out due to the gauge choice (\ref{eq: bdy_gauge_field}) (each azimuthal integral is of the form $\int_{-1}^{1} x \, dx =0 $) and so the mass charge becomes a purely gravitational issue.\footnote{Even if we used a different gauge for $A_{\mu}$, the analysis would go through in exactly the same manner as \cite{Papadimitriou:2005ii} and the Maxwell contribution to the Wald Hamiltonian would match the contribution to the holographic mass.} After working through all of the algebra one finds the bare Hamiltonian contributes 
\begin{equation} \label{eq: bare_mass_formula}
   \delta H_{\xi} = \lim_{\epsilon \rightarrow 0} \int_{\Sigma} \delta  \mathbf{Q} - i_{\xi} \bm{\Theta} = \lim_{\epsilon \rightarrow 0} \left[ \frac{1}{\epsilon \kappa}   \delta (\mu_+ + \mu_-) + \frac{1}{2 \kappa} (3 m \delta K + 2 K \delta m ) \right],
\end{equation}
where we regulate using the inverse radial coordinate defined in (\ref{eq: z_coord_def}) as $z=\epsilon>0$. We see that this term is clearly divergent due to the presence of the $\mathcal{O}(1/\epsilon)$ term in the asymptotic expansion on the right hand side and will need to be supplemented by the corner term in (\ref{eq: H_ren}). This term is constructed from the counterterm Lagrangian 
\begin{equation}
    \mathbf{L}_{\text{ct}} = - \frac{1}{16\pi G} \sqrt{-h} ( 4 + \mathcal{R} ) dt \wedge d\theta \wedge d\varphi
\end{equation}
and so the corner symplectic potential is just that of three dimensional Einstein gravity, up to an overall minus sign. The only formula we need is thus 
\begin{equation}
    \bm{\Theta}_{\text{ct}} = \frac{1}{16 \pi G}\cdot \frac{1}{2!} \varepsilon_{i j k} \left(D_{l} \delta h^{l i} - D^{i} (h_{ l m} \delta h^{l m} ) \right) dx^{j} \wedge dx^{k},
\end{equation}
where $\varepsilon_{t \theta \phi} = \sqrt{-h}$, $D$ is the Levi-Civita connection associated with $h$ and all indices are understood to be moved with $h$. This will be sufficient to compute the corner improvement in (\ref{eq: H_ren}). Explicitly, we have 
\begin{align}
\begin{split}
   \lim_{\epsilon \rightarrow 0} \int_{\partial \Sigma}  i_{\xi} \bm{\Theta}_{\text{ct}} & = \lim_{\epsilon \rightarrow 0} \left( \int_{S^1_+}  i_{\xi} \bm{\Theta}_{\text{ct}} -  \int_{S^1_-}  i_{\xi} \bm{\Theta}_{\text{ct}} \right) \\
   & = \lim_{\epsilon \rightarrow 0} \left\{ - \frac{1}{\epsilon \kappa} \delta (\mu_+ + \mu_-) - \frac{\alpha}{\kappa} \left[ \delta (\Xi K m \alpha ) + \Xi \alpha m \delta K \right] \right\},
       \end{split}
\end{align}
where we can immediately see that this term acts as an $\mathcal{O}(1/\epsilon)$ correction to the bare charge formula (\ref{eq: bare_mass_formula}). The finite term is a little less obvious but after some algebraic manipulation we can show that the fully renormalised Hamiltonian is
\begin{equation} \label{eq: H_ren_mass}
     \delta H^{\text{ren}}_{\xi} = \delta \mathcal{M}_{\text{hol}} + \frac{m}{2 \kappa^2} \left\{ 2 K \alpha \kappa \Xi \delta \alpha - 2 K (\alpha^2 \Xi -1) \delta \kappa - \kappa( 2\alpha^2 \Xi -1) \delta K \right\} = \delta \mathcal{M}_{\text{hol}},
\end{equation}
where in the second equality we used the well-posedness constraint (\ref{eq: variational_constraint_general}). This can trivially be integrated in phase space to prove that 
\begin{equation} \label{eq: mass_equivalence}
    H^{\text{ren}}_{\xi} = \mathcal{M}_{\text{hol}}  = \mathcal{M},
\end{equation}
concluding the proof of equivalence between the holographic mass formula and Wald Hamiltonian for the solution (\ref{eq: charged_acc_bh_metric})-(\ref{eq: charged_acc_bh_gauge_field}).
\subsection{Electric charge}

The next charge we will need to define is the electric charge, which can be done in a straightforward manner using (\ref{eq: electric_Hamiltonian}), taking $f$ to be a constant. We follow \cite{Papadimitriou:2005ii} in  computing the $U(1)$ Noether charge form as 
\begin{equation}
    \mathbf{Q}[f] = - \frac{1}{4 \pi G} f * \mathbf{F} 
\end{equation}
and the electric charge is defined as 
\begin{equation} \label{eq: electric_charge} 
 Q_e = H_{-1}  = \int_{\Sigma_{\infty}} \mathbf{Q}[-1] = \frac{1}{4 \pi G} \int_{\Sigma_{\infty}} * \mathbf{F} = \int_{\Sigma_{\infty}} d^2 x \sqrt{-g_{(0)}} j^t,
\end{equation}
where we assume w.l.o.g that $f \rightarrow -1$ on $\mathscr{I}$. This picks the opposite sign convention to \cite{Papadimitriou:2005ii} but will give the same value for $Q_e$ as we use a volume element with the opposite sign to that work. Our result matches equation (3.24) of \cite{Cassani:2021dwa} and thus completes the derivation of the electric charge from the covariant phase space. In order to evaluate this charge for the solution (\ref{eq: charged_acc_bh_gauge_field}), we compute explicitly
\begin{align} 
    \mathbf{F} & = g K \sin \theta  d\theta \wedge d\varphi -\frac{e }{\kappa  r^2}dt \wedge dr,\label{eq: F_bulk} \\
    *\mathbf{F} & = e K \sin \theta d \theta \wedge d \varphi +\frac{g}{\kappa  r^2} dt \wedge dr \label{eq: F_dual} 
\end{align}
and thus the electric charge is given by 
\begin{equation}
Q_e = \frac{1}{4\pi G} \int_{0}^{\pi} d\theta \int_{0}^{2\pi} d\varphi \, e K \sin \theta  = \frac{e K}{G}.
\end{equation} 

Looking ahead to the first law, we also note that the electric charge can be defined as an integral over a cross-section of the horizon $\Sigma_{\mathcal{H}}$ (as opposed to a cross-section of the conformal boundary $\Sigma_{\infty}$), which we can take w.l.o.g. to be the bifurcation 2-surface \cite{Wald:1993nt, Jacobson:1993vj}. To see this, we recall Maxwell's equations
\begin{equation}
    \mathbf{E}_{\text{M}} = d * \mathbf{F}
\end{equation}
and consider now a constant time slice $C$ which stretches between the black hole horizon and the conformal boundary. When on-shell we have 
\begin{equation}
    0 \approx \int_{C} d * \mathbf{F} = \int_{\partial C} * \mathbf{F} = \int_{\Sigma_{\infty}} * \mathbf{F} -  \int_{\Sigma_{\mathcal{H}}} * \mathbf{F} +  \int_{\mathcal{S}_-} * \mathbf{F} -  \int_{\mathcal{S}_+} * \mathbf{F},
\end{equation}
by Stokes' theorem. Using (\ref{eq: F_dual}) we see that $(*\mathbf{F})_{r\varphi}=0$ and thus the string terms do not contribute. This allows us to write the electric charge as an integral over the horizon
\begin{equation} \label{eq: Electric_horizon}
    Q_e = \frac{1}{4 \pi G} \int_{\Sigma_{\infty}} * \mathbf{F} =  \frac{1}{4 \pi G} \int_{\Sigma_{\mathcal{H}}} * \mathbf{F},
\end{equation}
a fact which will be crucial in our derivation of the first law.

\subsection{Magnetic charge} \label{sec: magnetic_charge}

Our final conserved charge is the magnetic charge $Q_m$, an object which seems somewhat difficult to define using the covariant phase space approach: we have already assigned conserved charges to both the time translation Killing vector $\partial_t$ and the constant $U(1)$ gauge transformation so it seems like there is nothing left to produce additional charges! (The axial Killing field $\partial_{\varphi}$ is associated to angular momentum, which vanishes trivially in the case we consider.) As it turns out, one can define the magnetic charge using the covariant phase space formalism by adding a \textit{topological term} to the action, the details of which we provide in appendix \ref{sec: Magnetic_charges}. This term will play no role in the first law, so we leave the detailed discussion of this term as an aside.

The magnetic charge is given by 
\begin{equation} \label{eq: Q_m}
    Q_m =  \frac{1}{4 \pi G} \int_{\Sigma_{\infty}} \mathbf{F} = \frac{g K}{G},
\end{equation}
where we used (\ref{eq: F_bulk}) to evaluate the charge explicitly. Following a similar line of logic to the electric charge, the magnetic charge can also be written as an integral over the bifurcation surface by using the Bianchi identity for the field strength tensor 
\begin{equation}
        0 = \int_{C} d  \mathbf{F} = \int_{\partial C}  \mathbf{F} = \int_{\Sigma_{\infty}}  \mathbf{F} -  \int_{\Sigma_{\mathcal{H}}}  \mathbf{F} +  \int_{\mathcal{S}_-}  \mathbf{F} -  \int_{\mathcal{S}_+}  \mathbf{F}
\end{equation}
and in much the same manner as the electric argument, the cosmic string terms do not contribute, leaving 
\begin{equation} \label{eq: Magnetic_horizon}
    Q_m =  \frac{1}{4 \pi G} \int_{\Sigma_{\infty}} \mathbf{F} =  \frac{1}{4 \pi G}  \int_{\Sigma_{\mathcal{H}}}  \mathbf{F}.
\end{equation}

\section{First law} \label{sec: first_law}

With all of the charges defined, we are almost ready to move on to the derivation of the first law. First we have to establish the definitions of the other important quantities in the law which are not explicit conserved charges. The first is the Bekenstein-Hawking entropy, given by the usual formula in terms of the horizon area $\mathcal{A}$
\begin{equation}
S_{\text{BH}} = \frac{\mathcal{A}}{4 G} = \frac{1}{4 G} \int_{0}^{\pi} d\theta \int_0^{2\pi} d\varphi \left.\sqrt{g_{\theta \theta}  g_{\varphi \varphi} }\right|_{r=r_+} = \frac{K \pi r_+^2 }{G(1-\alpha^2r_+^2)}.
\end{equation}

The second quantity is the black hole temperature $T$. We recall that this is defined as $T=\beta^{-1} = \frac{\kappa_{\text{sg}}}{2\pi}$, where the surface gravity $\kappa_{\text{sg}}$ is constructed from the horizon generator $\xi = \partial_t$ via $\kappa_{\text{sg}}^2 = -\frac{1}{2} \nabla_{\mu} \xi_{\nu} \nabla^{\mu} \xi^{\nu}$. Utilising these definitions, we find the temperature of the black hole to be 
\begin{equation} \label{eq: Temperature}
    T = \frac{Q'(r_+)}{4 \kappa \pi r_+^2}.
\end{equation}

We note consistency with \cite{Wald:1993nt, Iyer:1994ys} in that these objects can be constructed in terms of the gravitational part of the Noether charge form (\ref{eq: Q}) 
\begin{equation} \label{eq: Wald_entropy}
    \int_{\Sigma_{\mathcal{H}}} \mathbf{Q}_{\text{EH}}[\xi] = T S_{\text{BH}},
\end{equation}
a fact which will reappear in our derivation of the first law. 

The next quantities we need to define are the potentials dual to the electric and magnetic charges respectively \cite{Papadimitriou:2005ii, Cassani:2021dwa}. In the electric case we have the electrostatic potential defined via 
\begin{equation} \label{eq: Phi_e}
    \Phi_e \equiv \Phi_{\infty} - \Phi_{H} = - \Phi_{H} = - \left. i_{\xi} \mathbf{A} \right|_{r=r_+},
\end{equation}
where we used $\Phi_{\infty} = 0$ which is a result of the gauge choice (\ref{eq: bdy_gauge_field}) as there are no $\theta$-independent terms in $i_{\xi} \mathbf{A}_{(0)}$ \cite{Cassani:2021dwa}. Now applying equation (\ref{eq: charged_acc_bh_gauge_field}) allows us to read off the potential as 
\begin{equation}
    \Phi_e = \frac{e}{\kappa r_+ }.
\end{equation}

The magnetic potential is slightly more subtle in that in \cite{Cassani:2021dwa} it was simply introduced as the electric-magnetic dual of $\Phi_e$ by replacing $e \rightarrow g$. Here we will discuss how this can be realised as a potential dual to the magnetic charge. We note that we can write the magnetic charge as 
\begin{equation}
    Q_m = \frac{1}{4\pi G} \int_{\Sigma_{\infty}} *\mathbf{G},
\end{equation}
where $\mathbf{G} = - * \mathbf{F}$. In order to compute the magnetic potential, we first compute the dual gauge field $\tilde{\mathbf{A}}$ that sources $\mathbf{G}$, i.e. $d \tilde{\mathbf{A}} = \mathbf{G}$. Using (\ref{eq: F_dual}) we find 
\begin{equation} \label{eq: A_dual} 
    \tilde{\mathbf{A}} = -\frac{g}{r \kappa} dt + e K \cos \theta d \varphi,
\end{equation}
where we note that 
\begin{equation}
     \tilde{\mathbf{A}}_{(0)} = \lim_{\epsilon \rightarrow 0} \tilde{A}_i|_{z = \epsilon} dx^i  = \cos \theta \left( - \frac{g \alpha}{\kappa} dt + e K d\varphi \right)
\end{equation}
and we have again chosen a gauge s.t. $\int_{0}^{\pi} d\theta  \tilde{A}^{(0)}_i = 0$. Having established the magnetic gauge field, we can now define the magnetic potential as 
\begin{equation} \label{eq: Phi_m}
    \Phi_m = \tilde{\Phi}_{\infty} - \tilde{\Phi}_{H} = - \tilde{\Phi}_{H}  = -  \left. i_{\xi} \tilde{\mathbf{A}} \right|_{r=r_+} = \frac{g}{\kappa r_+},
\end{equation}
which we see can be obtained from $\Phi_e$ under replacement of $e \rightarrow g$.

\subsection{Geometric derivation}

With all of the charges $(\mathcal{M}, \, Q_e, \, Q_m)$ and auxiliary quantities $(S_{\text{BH}}, \, T, \, \Phi_e, \, \Phi_{m})$ defined, we are finally ready to derive the first law using the covariant phase space. In doing this, we follow \cite{Wald:1993nt, Iyer:1994ys} by considering a spacelike slice of spacetime $C$ which stretches from a cross-section of the black hole horizon $\Sigma_{\mathcal{H}}$ (which can be taken without loss of generality to be the bifurcation surface \cite{Jacobson:1993vj}) out to a cross-section of conformal infinity $\Sigma_{\infty}$. The novel aspect of this surface in case of the accelerating solution is that the boundary of $C$ does not merely consist of the two aforementioned surfaces but also includes the two cosmic strings responsible for the black hole acceleration! Technically speaking, we have
\begin{equation}
    \partial C = \Sigma_{\infty} - ( \Sigma_{\mathcal{H}} - \mathcal{S}_- + \mathcal{S}_+ ),
\end{equation}
where the signs are due to the induced orientations on each surface. A similar boundary structure has already been considered in the study of accelerating black hole thermodynamics in asymptotically (locally) flat spacetime \cite{Ball:2020vzo, Ball:2021xwt} and we will see that this will also play a crucial role in the thermodynamics of the AlAdS case. 

The derivation of the first law follows the same logic as \cite{Wald:1993nt}, namely in that we begin with an integral of the symplectic current over the surface $C$ 
\begin{equation} \label{eq: first_law_schematic}
   0  = \int_{C} \bm{\omega}[\psi; \delta \psi, \mathcal{L}_{\xi} \psi] = \int_{\partial C} \bm{k}_{\xi} = \int_{\Sigma_{\infty}} \bm{k}_{\xi} - \int_{\Sigma_{\mathcal{H}}} \bm{k}_{\xi} + \int_{\mathcal{S}_-} \bm{k}_{\xi} - \int_{\mathcal{S}_+} \bm{k}_{\xi},
\end{equation}
where $\xi$ is the horizon generator given in equation (\ref{eq: xi_killing}) and the integral vanishes by virtue of $\xi$ being a Killing vector. In the series of equalities on the right hand side above we have introduced
\begin{equation} \label{eq: k}
    \bm{k}_{\xi} = \delta \mathbf{Q}[\xi] - i_{\xi} \mathbf{\Theta}[\psi; \delta \psi] + d i_{\xi} \mathbf{\Theta}_{\text{ct}}[h_{ij}; \delta h_{ij}],
\end{equation}
which can be seen from equation (\ref{eq: omega_exact}) and we have chosen to add the additional $d$-exact term on the right hand side. We have already seen that the inclusion of this exact term gives the correct definition of the charges (\ref{eq: H_ren}) and it will also be an elegant choice in explaining the role of each boundary contribution in the first law. Note however that this term is not necessary as the first law is invariant under transformations of the form $ \bm{k}_{\xi} \rightarrow  \bm{k}_{\xi} + d\mathbf{B}$. This can be seen directly from (\ref{eq: first_law_schematic}) where the orientations mean that all corner contributions cancel. We now will provide the analysis of each term in (\ref{eq: first_law_schematic}) in order to derive the first law.

\subsubsection{Conformal boundary term}

The first term we will analyse is the contribution at the conformal boundary, namely 
\begin{equation}
    \int_{\Sigma_{\infty}} \bm{k}_{\xi} = \int_{\Sigma_{\infty}} \delta \mathbf{Q}[\xi] - i_{\xi} \mathbf{\Theta}[\psi; \delta \psi] + d i_{\xi} \mathbf{\Theta}_{\text{ct}}[h_{ij}; \delta h_{ij}].
\end{equation}
Upon using equations (\ref{eq: H_ren}) and (\ref{eq: H_ren_mass}) we immediately see
\begin{equation} \label{eq: law_mass}
     \int_{\Sigma_{\infty}} \bm{k}_{\xi} = \delta H_{\xi}^{\text{ren}} = \delta \mathcal{M},
\end{equation}
so the term at the conformal boundary contributes precisely the variation of the mass charge.

\subsubsection{Horizon term}

The horizon term is 
\begin{equation}
    \int_{\Sigma_{\mathcal{H}}} \bm{k}_{\xi} = \int_{\Sigma_{\mathcal{H}}} \delta \mathbf{Q} - i_{\xi} \mathbf{\Theta} 
\end{equation}
and to analyse this we will use the split of the symplectic quantities into their ``Einstein-Hilbert" and ``Maxwell'' components as in (\ref{eq: symp_split}).

The gravitational piece contributes 
\begin{equation}
     \int_{\Sigma_{\mathcal{H}}} \bm{k}^{\text{EH}}_{\xi} =  \int_{\Sigma_{\mathcal{H}}} \delta \mathbf{Q}_{\text{EH}} = \delta( T  S_{\text{BH}} ),
\end{equation}
where the fact that $\xi$ vanishes at the bifurcation surface removes the $i_{\xi} \bm{\Theta}_{\text{EH}}$ term and then the Wald entropy formula (\ref{eq: Wald_entropy}) ensures the right equality. This term simplifies further in that we follow \cite{Wald:1993nt, Papadimitriou:2005ii} in choosing perturbations such that we match the horizons of the perturbed and unperturbed solutions, as well as the unit surface gravity generators of the horizons $\tilde{\xi} = \frac{1}{\kappa_{\text{sg}}} \xi$. We immediately have $\delta \kappa_{\text{sg}} = 0$ (as $\delta \xi =0$), so this in turn leads to $\delta T= 0$. This allows us to write the final form of the horizon term as 
\begin{equation} \label{eq: Horizon_term_final}
 \int_{\Sigma_{\mathcal{H}}} \bm{k}^{\text{EH}}_{\xi}=  T \delta S_{\text{BH}},
\end{equation}
precisely as one would find for AlAdS black holes without cosmic string insertion \cite{Papadimitriou:2005ii}.

The electromagnetic piece is 
\begin{equation}
     \int_{\Sigma_{\mathcal{H}}} \bm{k}^{\text{M}}_{\xi} =  \int_{\Sigma_{\mathcal{H}}} \delta \mathbf{Q}_{\text{M}} - i_{\xi} \mathbf{\Theta}_{\text{M}} ,
\end{equation}
where we note that the $i_{\xi} \bm{\Theta}_{\text{M}}$ can no longer be ignored. We observe from (\ref{eq: Phi_e}) that $i_{\xi} \mathbf{A}|_{r=r_+}$ is non-zero and thus one needs to treat the contractions of the Maxwell symplectic potential and the horizon generator more carefully. We also note that the gauge field (\ref{eq: charged_acc_bh_gauge_field}) is not regular at the black hole horizon \cite{Ferrero:2020twa, Cassani:2021dwa}, a statement which is generically true for spacetimes with magnetic charge. Analysing this term more carefully we find
\begin{equation} \label{eq: omega_maxwell_1}
 \delta \mathbf{Q}_{\text{M}} - i_{\xi} \mathbf{\Theta}_{\text{M}} = - \frac{1}{4 \pi G} \left[ i_{\xi} \mathbf{A} \left( \delta * \mathbf{F} \right)  + \delta \mathbf{A} \wedge i_{\xi} * \mathbf{F} \right]
\end{equation}
and working on-shell so that $d * \mathbf{F} \approx 0$ and recalling that $\delta_{\xi} * \mathbf{F} = \mathcal{L}_{\xi} * \mathbf{F} = 0$, we have 
\begin{equation}
    0 = (i_{\xi} d + d i_{\xi} ) * \mathbf{F} \approx d i_{\xi} * \mathbf{F} \implies i_{\xi} * \mathbf{F} \approx d \mathbf{X} \qquad \text{(locally)}.  
\end{equation}
In order to solve for $\mathbf{X}$, we recall the dual field strength $d \tilde{\mathbf{A}} = \mathbf{G} = - * \mathbf{F}$ and using $\delta_{\xi} \tilde{\mathbf{A}} = 0$ we can write the equation above as 
\begin{equation}
i_{\xi} * \mathbf{F} = - i_{\xi} d \tilde{\mathbf{A}} = d i_{\xi} \tilde{\mathbf{A}} \approx d \mathbf{X} \implies i_{\xi} \tilde{\mathbf{A}} = \mathbf{X}.
\end{equation}
We apply this to equation (\ref{eq: omega_maxwell_1}) and use ``by parts" type manipulations in order to obtain
\begin{equation} \label{eq: k_Maxwell_final}
    \delta \mathbf{Q}_{\text{M}} - i_{\xi} \mathbf{\Theta}_{\text{M}} = - \frac{1}{4 \pi G} \left[  i_{\xi} \mathbf{A} \left( \delta * \mathbf{F} \right) + i_{\xi} \tilde{\mathbf{A}} \left( \delta \mathbf{F} \right) - d \left( \delta \mathbf{A} i_{\xi} \tilde{\mathbf{A}} \right)   \right],
\end{equation}
which we note is consistent with previous formulae derived in \cite{Lu:2013ura} and utilised in \cite{Ma:2022nwq}. When integrated over the bifurcation surface (\ref{eq: k_Maxwell_final}) gives
\begin{equation} \label{eq: horizon_em_terms}
      \int_{\Sigma_{\mathcal{H}}} \bm{k}^{\text{M}}_{\xi} = \Phi_e \delta Q_e + \Phi_m \delta Q_m,
\end{equation}
where we used equations (\ref{eq: Electric_horizon}), (\ref{eq: Magnetic_horizon}), (\ref{eq: Phi_e}), (\ref{eq: Phi_m}) and the fact that the corner integrals over the poles of the bifurcation surface cancel one another out. We note that due to this careful analysis, we have \textit{not} been forced to fix the electric and magnetic potentials between the perturbed and unperturbed solutions i.e. (\ref{eq: horizon_em_terms}) will still hold when $\delta \Phi_e \neq 0$ and $\delta \Phi_m \neq 0$. This is an advancement upon \cite{Papadimitriou:2005ii}, where the $i_{\xi} \bm{\Theta}$ term was not considered carefully enough at the horizon and they were forced to fix the value of the electric potential $\delta \Phi_e =0$. 

\subsubsection{Cosmic string terms}

The final contributions to the first law are those which are special to accelerating solutions, namely the thermodynamic length and tension terms which arise from the presence of the cosmic strings. We begin by recalling that the strings are located at $\theta_{-} = 0$ and $\theta_+ =\pi$ respectively and thus 
\begin{equation}
\int_{\mathcal{S_{\pm}}} \bm{k}_{\xi} = \lim_{\epsilon \rightarrow 0} \int_{r_+}^{\frac{1}{\epsilon \mp \alpha}} \int_{0}^{2\pi} \left. k_{r \varphi} \right|_{\theta_{\pm}} \, drd\varphi. 
\end{equation}
Using the general formula for the Maxwell contribution (\ref{eq: k_Maxwell_final}) we see that this term contributes nothing to the integral above except for the $d$-exact term which will not contribute to the first law. Thus we can treat the cosmic string terms as purely gravitational 
\begin{equation}
    \int_{\mathcal{S_{\pm}}} \bm{k}_{\xi} =   \int_{\mathcal{S_{\pm}}}  \delta \mathbf{Q}_{\text{EH}} - i_{\xi} \mathbf{\Theta}_{\text{EH}} + d i_{\xi} \mathbf{\Theta}_{\text{ct}}.
\end{equation}
Explicit computation using (\ref{eq: Q}) shows that the Noether charge 2-form does not contribute due to $\mathbf{Q} \stackrel{\mathcal{S_\pm}}{=} 0 $ and thus we have 
\begin{equation}
      \int_{\mathcal{S_{\pm}}} \bm{k}_{\xi} = \int_{\mathcal{S_{\pm}}}  - i_{\xi} \mathbf{\Theta}_{\text{EH}} + d i_{\xi} \mathbf{\Theta}_{\text{ct}} =  - \int_{\mathcal{S_{\pm}}}  i_{\xi} \mathbf{\Theta}_{\text{EH}} + \int_{S^1_{\pm}} i_{\xi} \mathbf{\Theta}_{\text{ct}},
\end{equation}
where we have used Stokes' theorem in order to separate the integral into bare and counterterm pieces just as we did for the mass charge. We will present these results of these terms one by one, illustrating again the elegance of the corner term. The bare contribution is 
\begin{equation} 
 \int_{\mathcal{S_{\pm}}} \bm{k}^{\text{bare}}_{\xi} =  - \int_{\mathcal{S_{\pm}}}  i_{\xi} \mathbf{\Theta}_{\text{EH}} = \mp \lambda^{\text{bare}}_{\pm} \delta \mu_{\pm},
\end{equation}
where 
\begin{equation} \label{eq: k_bare}
    \lambda^{\text{bare}}_{\pm} = - \frac{1}{ \kappa \epsilon} + \frac{r_+}{\kappa(1 \pm \alpha r_+)}
\end{equation}
are the \textit{bare thermodynamic lengths} and $\mu_{\pm}$ are the string tensions defined in (\ref{eq: cosmic_string_tensions}). This term clearly diverges as $\epsilon \rightarrow 0$. 

The counterterm contribution is 
\begin{equation}
 \int_{\mathcal{S_{\pm}}} \bm{k}^{\text{ct}}_{\xi} = \int_{S^1_{\pm}} i_{\xi} \mathbf{\Theta}_{\text{ct}} =  \mp \frac{1}{\kappa} \left[\frac{1}{\epsilon} - \alpha( 2 \alpha m  \pm \Xi) \right] \delta \mu_{\pm}
 \end{equation}
 and thus combining this term appropriately with (\ref{eq: k_bare}), we arrive at 
 \begin{equation} \label{eq: law_lengths}
     \int_{\mathcal{S_{\pm}}} \bm{k}_{\xi} = \mp \lambda_{\pm} \delta \mu_{\pm},
 \end{equation}
 where the \textit{renormalised thermodynamic lengths} are defined as
 \begin{equation}
     \lambda_{\pm} = \frac{r_+}{\kappa(1 \pm \alpha r_+)} - \frac{\alpha}{\kappa} (2 \alpha m  \pm \Xi ) = \frac{r_+}{\kappa(1 \pm \alpha r_+)} \mp \frac{\alpha}{\kappa} P_{\pm}.
 \end{equation}
 
\subsection{Final statement of the law}

Combining equations (\ref{eq: law_mass}), (\ref{eq: Horizon_term_final}), (\ref{eq: horizon_em_terms}) and (\ref{eq: law_lengths}) with the signs given in (\ref{eq: first_law_schematic}) we arrive at the final form of the first law 
\begin{equation} \label{eq: final_first_law}
 T \delta S_{\text{BH}} = \delta \mathcal{M} - \Phi_e \delta Q_e - \Phi_m \delta Q_m + \lambda_- \delta \mu_- + \lambda_+ \delta \mu_+,
\end{equation}
where, to recap, the relevant quantities are defined as  
\begin{equation}
\begin{alignedat}{2} \label{eq: all_physical_quantities} 
      \mathcal{M} & = \frac{K m (1 - \alpha^2 \Xi )}{\kappa G}, \\
       S_{\text{BH}} & = \frac{\mathcal{A}}{4G} = \frac{K \pi r_+^2 }{G(1-\alpha^2r_+^2)},  \qquad  && T  = \frac{Q'(r_+)}{4 \kappa \pi r_+^2}, \\ 
       Q_e  & = \frac{e K}{G}, \qquad && \Phi_e  = \frac{e}{\kappa r_+}, \\
        Q_m & = \frac{g K}{G}, \qquad &&   \Phi_m = \frac{g}{\kappa r_+},\\
        \mu_{\pm} & = \frac{1}{4 G} \left ( 1 -P_{\pm} K \right), \qquad  && \lambda_{\pm}   = \frac{r_+}{\kappa(1\pm \alpha r_+)} \mp \frac{\alpha}{\kappa} P_{\pm}.
\end{alignedat}
\end{equation}
We note that all of these quantities are identical to those of \cite{Cassani:2021dwa}, \textit{except} $\lambda_\pm$ which are different because our phase space of parameters is different and not due to the fact that we derived our laws using different techniques. We demonstrate this by deriving the first law (\ref{eq: final_first_law}) using the ``horizon polynomial'' method of \cite{Appels:2016uha} in Appendix \ref{sec: horizon_poly}, where we also demonstrate explicitly the reasons for the differences in the thermodynamic length terms. The law presented above can be seen as a five-parameter law: one starts with the six parameters $\{m, e, g, \alpha, K, \kappa\}$ and then the constraint equation (\ref{eq: variational_constraint_general}) reduces this by one.

It is important to discuss some of the key differences with the law presented in \cite{Cassani:2021dwa} and ours: the law presented in equation (3.46) of \cite{Cassani:2021dwa} is a full cohomogeneity law with an ill-posed variational problem whereas (\ref{eq: final_first_law}) is a cohomogeneity-1 law with a well-posed variational problem. In fact, the variations that enter the first laws of \cite{Appels:2016uha, Appels:2017xoe, Gregory:2017ogk, Anabalon:2018ydc, Anabalon:2018qfv, Cassani:2021dwa} (when $\delta \mu_{\pm} \neq 0$) are all generically \textit{ill-posed} and this results in their expressions for the thermodynamic lengths $\lambda_{\pm}$ differing from ours. The crucial feature of well-posedness is in demonstrating the equivalence of the holographic mass and the Wald Hamiltonian, as was shown in equation (\ref{eq: mass_equivalence}). Due to this equivalence, our first law (\ref{eq: final_first_law}) can be taken to be read with either quantity acting as $\mathcal{M}$ in the law and is thus entirely consistent. This is in great contrast to \cite{Cassani:2021dwa} where the holographic mass is not equivalent to the Wald Hamiltonian and thus the first law changes form depending on the charge that appears in the law. As explained in detail in Appendix \ref{sec: differences}, if one writes their first law with the Wald Hamiltonian $H_{\xi}$ then one finds the same $\lambda_{\pm}$ as given in (\ref{eq: all_physical_quantities}). If one uses the holographic mass $\mathcal{M}_{\text{hol}}$ then one finds the $\lambda_{\pm}$ as given in equation (3.43) of \cite{Cassani:2021dwa}. This inconsistency in the form of the law is manifested due to ill-posedness and thus we strongly advocate a first law where the variations satisfy the well-posedness constraint (\ref{eq: variational_constraint_general}). 

There are various choices of $\kappa$ which solve the master equation (\ref{eq: variational_constraint_general}) that one may now want to examine. One obvious choice is to take $\kappa$ as a phase space constant such that $\delta \kappa = 0$. This is the clearest limit from the perspective of both Einstein-Maxwell theory in four dimensions and the dual CFT$_3$: all of the parameters $\{ m, e, g, \alpha, K \}$ are physically well understood, and the fact that $\kappa$ is fixed on phase space means that different choices correspond to scaling of dimensionful quantities in the dual field theory \cite{Cassani:2021dwa}. On the other hand, allowing for a phase space dependent $\kappa$ may yet be crucial to discuss the thermodynamics of the supersymmetric solutions \cite{Klemm:2013eca} and thus the uplift into $d=11$ supergravity \cite{Ferrero:2020twa}. We shall now show that a phase space dependent $\kappa$ is important in consistently analysing the thermodynamics of a class of solutions that we shall define as \textit{close-to-supersymmetric and close-to-extremal} spindles.   

\section{Spindles and supersymmetry} \label{sec: spindles_susy}

In this section we will discuss various relations between the conserved charges $\{\mathcal{M}, Q_e, Q_m\}$ as given in (\ref{eq: all_physical_quantities})  and  applications of the first law (\ref{eq: final_first_law}) when we constrain the parameters of the solution. These constraints will arise by requiring various combinations of supersymmetry, extremality, and for the surfaces of constant $(t,r)$ to have the topology of a \textit{spindle}. We will begin with the technical details of these requirements, before applying them to derive loci satisfied by the conserved charges and a number of applications of the first law.

\subsection{Overview}

We have so far kept the choice of the deficit parameter $K$ entirely generic and thus considered a setup with conical singularities at both poles due to the presence of two cosmic strings. Within this class of solutions, a particularly interesting case is that of the constant $(t, r)$ surfaces having the topology of a \textit{spindle}. Following \cite{Ferrero:2020twa, Cassani:2021dwa}, we note that such a topology is obtained when we choose
\begin{equation} \label{eq: Spindle_period}
K = \frac{1}{n_+ P_+} = \frac{1}{n_- P_-},
\end{equation}
where $n_{\pm}$ are \textit{coprime positive integers} i.e. $\text{gcd}(n_-, n_+) =1$. With this choice, one has $\Sigma \cong \mathbb{WCP}^1_{[n_-, n_+]}$, the orbifold space known as a spindle. Such objects have been the topic of much recent study in the supergravity context \cite{Cassani:2021dwa, Ferrero:2020laf, Ferrero:2020twa, Ferrero:2021wvk}
due to their remarkable property that despite being singular surfaces (and thus inducing conical singularities when present in the low-dimensional spacetime) they admit smooth solutions when suitably uplifted into $d=11$ supergravity. When working with a spindle we can rewrite the cosmic string tensions $\mu_{\pm}$ entirely in terms of the spindle data, namely the coprime integers $n_{\pm}$. Using equation (\ref{eq: cosmic_string_tensions}) together with (\ref{eq: Spindle_period}) we find 
\begin{equation} \label{eq: spindle_tensions}
    \frac{1}{n_{\pm}} = 1 - 4 G \mu_{\pm}
\end{equation}
and we also note that the orbifold Euler characteristic of $\Sigma$ is given by \cite{Cassani:2021dwa}
\begin{equation} \label{eq: Euler_char}
    \chi = \frac{1}{n_-}+\frac{1}{n_+} = 2 - 4G (\mu_- + \mu_+)
\end{equation}
and is thus determined purely by the overall conical deficit (\ref{eq: overall_deficit}) present in the spacetime.

A crucial property for the uplift into regular solutions in supergravity is the property that the four dimensional solutions (\ref{eq: charged_acc_bh_metric})-(\ref{eq: charged_acc_bh_gauge_field}) are supersymmetric, i.e. that there exists a solution to the Killing spinor equation. Such equations have been analysed in \cite{Klemm:2013eca, Ferrero:2020twa, Cassani:2021dwa} and result in the following constraints between the parameters 
\begin{align} 
    g & = \alpha m, \label{eq: susy_constraints_1} \\
    g^2 & = \Xi ( \Xi -1 ). \label{eq: susy_constraints_2}
\end{align}
A supersymmetric solution is also \textit{extremal} ($T=0$) if \cite{Ferrero:2020twa}
\begin{equation} \label{eq: extremality}
e= 0 \implies Q_e = 0 . 
\end{equation}

\subsection{Charge loci}

It is natural to ask what algebraic constraints are placed upon the conserved charges $\{ \mathcal{M}, \, Q_e, \, Q_m\}$ given in (\ref{eq: all_physical_quantities}) when one applies various combinations of the supersymmetry, extremality and spindle topology conditions. We will now examine some interesting combinations and derive the various charge loci that result.

\subsubsection{Supersymmetric locus}

The first case of interest is to apply the supersymmetry constraints (\ref{eq: susy_constraints_1})-(\ref{eq: susy_constraints_2}) in order to write down the supersymmetric locus of charges. We note that supersymmetric solutions no longer have a black hole horizon\footnote{They exhibit a naked curvature singularity, visible at the conformal boundary $\mathscr{I}$.} \cite{Ferrero:2020twa}, although they can still be slowly accelerating and exhibit a single conformal boundary with representative (\ref{eq: bdy_metric_charged}) and energy-momentum tensor (\ref{eq: T_charged}). It is for these reasons that our analysis of the conserved charges should extend without issue to this class of solutions. In order to derive the locus, we find it helpful to rewrite (\ref{eq: susy_constraints_1})-(\ref{eq: susy_constraints_2}) as
\begin{equation} \label{eq: parameters_new}
    m = \frac{g}{\alpha}, \qquad \alpha^2 \Xi = \frac{g^2}{e^2 + g^2 } = \frac{Q_m^2}{Q_e^2 + Q_m^2}
\end{equation}
and thus we find the supersymmetric locus of charges is given by 
\begin{equation} \label{eq: susy_locus}
    \mathcal{M}  = \frac{Q_m Q_e^2}{\alpha \kappa (Q_e^2 + Q_m^2)}.
\end{equation}

\subsubsection{Supersymmetric and extremal locus}

An immediate application of the supersymmetry locus (\ref{eq: susy_locus}) is to the \textit{supersymmetric and extremal} black holes. Application of the extremality constraint (\ref{eq: extremality}) immediately results in 
\begin{equation} \label{eq: susy_extremal_locus}
    Q_e = \mathcal{M} = 0.
\end{equation}

This result seems somewhat surprising at first but these are still genuine black hole solutions as discussed in \cite{Ferrero:2020twa}. The vanishing of $\mathcal{M}$ imposes no further constraints upon the parameters than those already given in (\ref{eq: susy_constraints_1})-(\ref{eq: extremality}), in particular one still has $m, g, \alpha, K \neq 0$ and a highly non-trivial global structure, including the presence of acceleration horizons \cite{Ferrero:2020twa}. It is important to note that these acceleration horizons split the conformal boundary into two pieces \cite{Ferrero:2020twa} and thus it is unclear if (\ref{eq: susy_extremal_locus}) is really a relationship between the true conserved charges of the supersymmetric and extremal solution. It would be interesting to investigate this issue more deeply in future work. 

\subsubsection{Supersymmetric spindle locus}

We note that (\ref{eq: susy_locus}) is quantitatively different from the $a \rightarrow 0$ limit of the rotating supersymmetric locus as given in equation (1.1) of \cite{Cassani:2021dwa}. This is because we have applied the supersymmetry conditions \textit{before} fixing the solution to have the topology of a spindle. In order to work explicitly with a spindle, we start as in \cite{Cassani:2021dwa} by requiring that we fix the spindle data (\ref{eq: spindle_tensions}), namely we require
\begin{equation} \label{eq: fixed_spindle}
 \delta n_{\pm} = 0 \iff  \delta \mu_{\pm} = 0,
\end{equation}
where we used equation (\ref{eq: spindle_tensions}) in writing down the iff statement. Using equations (\ref{eq: overall_tension}) and (\ref{eq: overall_deficit}) we see that this corresponds to fixing the products $\alpha m K$ and $\Xi K$ which can be physically understood as fixing the overall tension and deficit respectively. Fixing the overall deficit has an important implication for the well-posedness constraint (\ref{eq: variational_constraint_general}): it is equivalent to equation (\ref{eq: kappa_old_constraint}) and thus in order to ensure well-posedness we must fix $\kappa$ as in equation (\ref{eq: kappa_old}), namely 
\begin{equation} \label{eq: kappa_spindle}
    \kappa = \sqrt{\Xi(1-\alpha^2 \Xi)}.
\end{equation}
This is consistent with the observation made in \cite{Anabalon:2018ydc, Anabalon:2018qfv, Cassani:2021dwa} that when one fixes the string tensions the choice of $\kappa$ given above ensures well-posedness. Using (\ref{eq: parameters_new}), we find
\begin{equation} 
    \kappa = \frac{Q_e Q_m}{\alpha (Q_e^2 + Q_m^2)}
\end{equation}
and thus the supersymmetric locus for a spindle is
\begin{equation}
    \mathcal{M} = Q_e.
\end{equation}
We note that fixing a spindle is actually a stronger than necessary requirement in order to arrive at this locus. The locus would be equivalent if one merely demands supersymmetry (\ref{eq: susy_constraints_1})-(\ref{eq: susy_constraints_2}) and a fixed Euler characteristic (\ref{eq: Euler_char}) (which is equivalent to fixing the overall deficit (\ref{eq: overall_deficit})). The supersymmetric and extremal spindle exhibits the same locus as (\ref{eq: susy_extremal_locus}).

\subsection{Applications of the first law}

We would like to apply our first law of slowly accelerating black hole thermodynamics (\ref{eq: final_first_law}) to the supersymmetric spindle solutions, although this is made difficult due to the global nature of such solutions. As mentioned, the supersymmetric and extremal solutions are not slowly accelerating and are thus beyond the scope of our first law. The next possible case of interest is the supersymmetric and non-extremal solutions (i.e. those where both (\ref{eq: susy_constraints_1}) and (\ref{eq: susy_constraints_2}) are satisfied but (\ref{eq: extremality}) is not) but as was shown in \cite{Ferrero:2020twa}, these do not even possess a black hole horizon and thus these are also entirely unsuitable! We will instead first apply our law to solutions with a fixed overall deficit angle, then those with a spindle as the constant $(t,r)$ surfaces. Finally, we will show that our law can be further restricted to the families of \textit{close-to-supersymmetric} and \textit{close-to-supersymmetric and close-to-extremal} spindles, which we shall introduce using criteria inspired by those given in Section 6.3 of \cite{Ferrero:2020twa}. We note that such solutions were referred to as ``non-supersymmetric and non-extremal" in \cite{Ferrero:2020twa} and we will alter this terminology here to avoid confusion with the other cases we study, none of which are sypersymmetric or extremal.

\subsubsection{Fixed deficit}

The first case we consider is to fix the overall deficit (\ref{eq: overall_deficit}) (equivalently the orbifold Euler characteristic (\ref{eq: Euler_char})) 
\begin{equation}
    \delta (\mu_+ + \mu_- ) =  \delta \chi = 0,
\end{equation}
a condition which results in equation (\ref{eq: kappa_spindle}) for well-posedness.
Applying this relation to the first law (\ref{eq: final_first_law}) we find the following reduction of the law 
\begin{equation} \label{eq: first_law_fixed_deficit}
     T \delta S_{\text{BH}} = \delta \mathcal{M} - \Phi_e \delta Q_e - \Phi_m \delta Q_m + (\lambda_+ - \lambda_-) \delta \mu_+,
\end{equation}
which is now a four-parameter law.

\subsubsection{Fixed spindle}

In order to write down the first law for a spindle, we start by applying equations (\ref{eq: fixed_spindle}) and (\ref{eq: kappa_spindle}) in order to fix the spindle topology and ensure well-posedness. We note that fixing the spindle topology corresponds to fixing the overall deficit (\ref{eq: overall_deficit}) and the overall tension (\ref{eq: overall_tension}) and thus can be obtained directly from (\ref{eq: first_law_fixed_deficit}) by imposing $\delta \mu_+ =0$. We can then immediately write down the first law for a spindle 
\begin{equation} \label{eq: first_law_non_susy_spindle}
     T \delta S_{\text{BH}} = \delta \mathcal{M} - \Phi_e \delta Q_e - \Phi_m \delta Q_m,
\end{equation}
which equivalently follows from (\ref{eq: final_first_law}) after fixing the string tensions and is a three-parameter law. 

\subsubsection{Close-to-supersymmetric spindle}

The next case is to apply some supersymmetry condition on top of (\ref{eq: fixed_spindle}). As we cannot apply both supersymmetry conditions, we follow Section 6.3 of  \cite{Ferrero:2020twa} in only applying 
\begin{equation} \label{eq: 1-BPS}
    g = \alpha m, 
\end{equation}
an equation which we take with  (\ref{eq: fixed_spindle}) to define a \textit{close-to-supersymmetric} spindle solution. The application of this constraint is straightforward in that we have 
\begin{equation}
       0 =  \delta ( \mu_- - \mu_+ ) = \frac{1}{G} \delta (\alpha m K) =\frac{1}{G}  \delta (g K) = \delta Q_m
\end{equation}
and so this just amounts to fixing the magnetic charge, reducing the first law of (\ref{eq: first_law_non_susy_spindle}) down to the ``standard" form of
\begin{equation} \label{eq: standard_first_law}
T \delta S_{\text{BH}} = \delta \mathcal{M} - \Phi_e \delta Q_e.
\end{equation}
We note that this is equivalent to that written down in \cite{Cassani:2021dwa} (with $J=0$) but is now a two-parameter law. 

\subsubsection{Close-to-supersymmetric and close-to-extremal spindle}

The final case of interest is the \textit{close-to-supersymmetric and close-to-extremal} spindle solution, which in addition to (\ref{eq: fixed_spindle}) and (\ref{eq: 1-BPS}) also has
\begin{equation} \label{eq: e=0}
e=  Q_e = 0 . 
\end{equation}
Note that this solution is not extremal as this condition only corresponds to extremality  when the solution is supersymmetric, i.e. when \textit{both} supersymmetry conditions (\ref{eq: susy_constraints_1})-(\ref{eq: susy_constraints_2}) hold. This solution is now characterised by a single parameter, which we can take to be the magnetic charge parameter $g$. In order to apply our first law to such a solution we need to specify the range of $g$ for which the solution is slowly accelerating. Such a range was computed in \cite{Ferrero:2020twa} and reads 
\begin{equation}
    g < g_{\text{BPS}} = \frac{\sqrt{1-\alpha^2}}{\alpha^2},
\end{equation}
where $g_{\text{BPS}}$ corresponds to the supersymmetric and extremal solution. Restricting $g$ to the range above, the first law for this solution is 
\begin{equation}
    T \delta S_{\text{BH}} = \delta \mathcal{M},
\end{equation}
which completes the application of our current law to slowly accelerating spindle solutions. We note that this reduction would not have been possible if $\kappa$ was not treated as a phase space dependent parameter. If $\kappa$ was taken to be a phase space constant, the master equation (\ref{eq: variational_constraint_general}) would reduce to a non-trivial constraint, and subsequently applying (\ref{eq: fixed_spindle}), (\ref{eq: 1-BPS}) and (\ref{eq: e=0}) would overconstrain the first law (\ref{eq: final_first_law}) down to a trivial (0-parameter) statement. This application thus provides important motivation to treat $\kappa$ as a generic parameter of the solution.
 
 It will be of future interest to study more carefully the effect of rotation: as well as adding an additional parameter, rotation also allows for slowly accelerating supersymmetric and extremal solutions \cite{Cassani:2021dwa}. This would open up many more interesting reductions of the more general first law, although the key issue of extracting the true mass for accelerating solutions must first be tackled. We leave this work to future studies.

\section{Conclusions and outlook} \label{sec: conclusions}

In this work we have extended the techniques of \cite{Papadimitriou:2005ii} in describing the charges and thermodynamics of AlAdS solutions to encompass \textit{accelerating} solutions in $d=4$ spacetime dimensions. This effort relied on two important developments, firstly the relaxation of Dirichlet boundary conditions to the more general demand that the variational problem is well-posed (\ref{eq: variational_principle_well_posed}). Secondly, one needs to supplement the definition of the Wald Hamiltonians by a suitable \textit{corner improvement} due the topology of $\mathscr{I}$: a hypersurface which is not smooth due to the presence of cosmic strings piercing the poles of the constant time cross-sections. With these improvements in place, we were able to show agreement for the conserved charges between the usual holographic definitions \cite{Papadimitriou:2005ii, Hollands:2005wt} and the charges constructed via the covariant phase space \cite{Wald:1993nt}.

The main motivation in developing these techniques was to examine the first law of thermodynamics for these solutions, and in particular to make comparison and contrast with the results of \cite{Appels:2016uha, Appels:2017xoe, Gregory:2017ogk, Anabalon:2018ydc, Anabalon:2018qfv, Cassani:2021dwa}. Of crucial importance in these works was the choice of the time scaling parameter $\kappa$ of which we have now shed more light on: we have shown that one can use the techniques developed in this paper to derive an entirely consistent first law of thermodynamics with $\kappa$ as long as one imposes the constraint (\ref{eq: variational_constraint_general}). We emphasise that the key idea in deriving this constraint is the requirement of a well-posed variational problem (\ref{eq: variational_principle_well_posed}), an idea which was previously considered for the restricted case of fixed string tensions in \cite{Anabalon:2018qfv}. However, such a requirement was not in place for the variations which change the overall deficit of the accelerating solution considered in \cite{Appels:2016uha, Appels:2017xoe, Gregory:2017ogk, Anabalon:2018ydc, Anabalon:2018qfv, Cassani:2021dwa} and this results in a mismatch between the holographic mass and the Wald Hamiltonian associated with the time translation. This mismatch results in an ambiguous form of the first law depending upon the ``mass" that appears in the law, manifested explicitly in finding different values for the thermodynamic lengths $\lambda_{\pm}$ for the holographic and Wald masses. In contrast, our first law (\ref{eq: all_physical_quantities}) is valid for both the holographic and Wald masses (because they are equivalent) and thus gives the true expression for the thermodynamic lengths $\lambda_{\pm}$. We note that when one fixes the overall deficit angle in the spacetime and one chooses the value of $\kappa$ as in (\ref{eq: kappa_old}) (as considered in \cite{Appels:2016uha, Appels:2017xoe, Gregory:2017ogk, Anabalon:2018ydc, Anabalon:2018qfv, Cassani:2021dwa}), the well-posedness constraint (\ref{eq: variational_principle_well_posed}) \textit{is} solved. This gives a concrete reason for the value of $\kappa$ used in \cite{Cassani:2021dwa} (the authors noted at the time that $\kappa$ ``gives the first law by trial and error''). 

One peculiar feature that arises in our approach is that unlike \cite{Appels:2016uha, Appels:2017xoe, Gregory:2017ogk, Anabalon:2018ydc, Anabalon:2018qfv, Cassani:2021dwa}, the master equation (\ref{eq: variational_constraint_general}) is not integrated on phase space (due to the non-integrability) and thus does not provide a closed form expression for $\kappa$ in terms of the other parameters. Equation (\ref{eq: variational_constraint_general}) can be understood as a constraint on variations of $\kappa$ and is crucial in the derivation of the first law (\ref{eq: final_first_law}). As a conseqeunce of (\ref{eq: variational_constraint_general}), the law effectively depends only on the variations of the (physically well-understood) parameters $m, e, g, \alpha, K$. We note however, due to their dependence on $\kappa$, one does not generically have (and cannot have due to non-integrability) closed expressions for the thermodynamic quantities $\mathcal{M}, T, \Phi_e, \Phi_m, \lambda_{\pm}$ (defined in (\ref{eq: all_physical_quantities})) in terms of $m, e, g, \alpha, K$.

This work thus provides a platform to study the thermodynamics of spindle solutions from the covariant phase space, but there are still many future directions which need to be pursued for a more complete understanding. The first issue is that of rotation, in particular in that one needs to first establish the true  mass charge for accelerating, rotating solutions. The difficulty in doing this lies in the fact that the true mass charge is associated to the timelike Killing vector of the solution when the boundary is in a \textit{non-rotating frame}, an issue which was resolved for AdS-Kerr-Newman black holes in \cite{Gibbons:2004ai, Papadimitriou:2005ii} but is yet to be entirely settled for the accelerating solutions. In \cite{Anabalon:2018qfv, Ferrero:2020twa, Cassani:2021dwa}, the mass charge is associated with the timelike Killing vector in a \textit{rotating} frame, and only matches the true mass along the poles of the spindle ($\theta =\theta_{\pm}$). We believe that a more careful analysis of the mass charge must first be carried out, with the $\Lambda$-BMS gauge fixing of \cite{Compere:2019bua} providing a possible algorithm in order to find the correct non-rotating frame. It would be interesting to observe how this would affect the first law, and would also open the door to apply the law to supersymmetric and extremal spindle solutions, where the rotation can be tuned so as to ensure slow acceleration \cite{Cassani:2021dwa}. 

In addition to rotation, it would also be interesting to include the other parameters of the Plebanski-Demia\'nski family as spelled out in \cite{Podolsky:2022xxd}. One could for example promote to thermodynamic objects the cosmological constant $\ell$ \cite{Caldarelli:1999xj, Kastor:2009wy, Cvetic:2010jb, Kubiznak:2015bya, Kubiznak:2016qmn} or Newton constant $G$ \cite{Cong:2021fnf, Visser:2021eqk}, the former of which may be understood in terms of a braneworld interpretation as recently discussed in \cite{Frassino:2022zaz}. Perhaps more interesting would be to include a non-zero NUT parameter $N$\footnote{This must be performed together with non-zero ``Kerr-like" rotation, as otherwise the accelerating, NUT-charged black holes fall outside the Plebanski-Demia\'nski family \cite{Podolsky:2020xkf, Astorino:2023elf, Astorino:2023ifg}.}. The thermodynamics of this parameter have been the subject of much recent study \cite{Hennigar:2019ive, Bordo:2019slw, Bordo:2019tyh, BallonBordo:2019vrn, Wu:2019pzr, Frodden:2021ces, Godazgar:2022jxm, Wu:2022rmx, Wu:2022mlz, Wu:2023woq, Liu:2022wku} and we note that the conical singularity structure we dealt with in this work via the covariant phase space is remarkably similar to that of the Taub-NUT-AdS as discussed in \cite{Godazgar:2022jxm}, with the difference being that the NUT charge introduces \textit{Misner strings} rather than cosmic strings along the poles of the constant $(t,r)$ surfaces. It would be interesting to combine these techniques to derive the thermodynamics for the entire class of solutions \cite{Podolsky:2022xxd}. It may also be of interest to understand the role of the NUT parameter in accelerating solutions where the underlying theory is something other than Einstein-Maxwell. As an example of this: accelerating, NUT charged black holes have recently been found in the theory of Einstein gravity conformally coupled to a scalar field (for $\Lambda =0$) \cite{Barrientos:2023tqb}. If such solutions are found in the $\Lambda < 0$ case, then the methods discussed in this paper may prove fruitful in analysing their charges and first law.

Another important future direction would be to move away from the slowly accelerating case and apply these techniques to AlAdS solutions with an acceleration horizon. In the asymptotically (locally) flat setting, the lack of a cosmological constant forces the inclusion of an acceleration horizon and covariant phase space techniques have been used \cite{Ball:2020vzo, Ball:2021xwt} to derive a first law using the background subtraction of the massless cosmic string spacetime. Such techniques should be readily applicable to the AlAdS case, and in fact should go further as the holographic counterterms will negate the need for background subtraction. Physically speaking, spacetimes containing multiple horizons with different surface gravities makes the assignment of a thermodynamic temperature unclear \cite{Kubiznak:2016qmn} and thus this issue would also have to be explored more deeply in the rapidly accelerating case. A direct application of this would be the \textit{supersymmteric and extremal} non-rotating solutions, as these must contain (for certain values of the azimuthal coordinate $\theta$) acceleration horizons \cite{Ferrero:2020twa}. These solutions possess a near-horizon geometry of AdS$_2 \times \mathbb{WCP}^1_{[n_-, n_+]}$ and thus are an important direction in understanding AdS$_2$ solutions, both from the lower dimensional and uplifted perspectives \cite{Boido:2022iye}.

Finally, we note that the techniques pioneered here should be applicable to spacetimes whose conformal boundary cross-sections $\Sigma_{\infty}$ are generic 2-manifolds-with-boundary as (\ref{eq: H_ren}) will still hold. We conjecture that this formula should apply immediately to higher even dimensions whereas odd dimensions may be more subtle due to the different boundary conditions one imposes due to the conformal anomaly \cite{Papadimitriou:2005ii}. While it is proven that there is no analogue of the C-metric in higher dimensions \cite{Podolsky:2006du} (and thus no good candidate solution to describing accelerating spacetime) such formulae may still be crucial in describing the charges and thermodynamics of more exotic solutions to the field equations.

\acknowledgments

This work is supported by the National Research Foundation of Korea under the grants, NRF-2022R1A2B5B02002247 (H.K, N.K, Y.L, and A.P.) and NRF-2020R1A2C1008497 (H.K. and A.P.). This research was partially supported by the Asia Pacific Center for Theoretical Physics (APCTP) via H.K. and A.P. participating in the APCTP SAG workshop, ``Entanglement, Large N and Black hole". A.P. would like to thank Matthew Roberts and Finn Larsen for insightful comments given during the workshop. We would like to thank an anonymous referee for their feedback and suggested improvements to the paper.

 \appendix

\section{Magnetic charges from the covariant phase space} \label{sec: Magnetic_charges}

In this appendix we show that the magnetic charge (\ref{eq: Q_m}) can be realised as a ture conserved charge in the covariant phase space formalism by adding an additional term to the action (\ref{eq: S_reg}). This term will be purely topological and thus will not contribute to the equations of motion, but will modify the symplectic structure (and thus the charges). We will show that this term does not affect either the variational problem (as studied in Section \ref{sec: variational_principle}) or the derivation of the first law (Section \ref{sec: first_law}) and thus the analysis of the main text is entirely sufficient to discuss the first law.

\subsection{Topological term}

As discussed in Section \ref{sec: magnetic_charge}, the standard discussion of Maxwell theory does not derive the magnetic charge as a conserved quantity from the covariant phase space point of view. In order to see how the magnetic charge is constructed on phase space we follow \cite{Godazgar:2022jxm} by considering the action (\ref{eq: S_reg}) supplemented by an additional term 
\begin{equation} 
S = S_{\text{bulk}} + S_{\text{bdy}} + S_{\text{top}},
\end{equation}
where the new term is
\begin{equation} \label{eq: S_top}
    S_{\text{top}} = - \frac{1}{8 \pi G} \int_{M} \mathbf{F}\wedge \mathbf{F}. 
\end{equation}
This term is often referred to as the ``$\theta$-term" \cite{Jackiw:1976pf, Callan:1976je} when added to the usual Yang-Mills Lagrangian and is responsible for the breaking of CP-symmetry in the quantum theory when included. As we will treat the  theory purely classically we will not worry about such features and we use it purely to illustrate this feature of the magnetic charge. 

The contribution from this new term to the equation of motion is trivial ($\mathbf{E}_{\text{top}} \equiv 0$) due to the closed field strength tensor $d \mathbf{F} = 0$ and thus we see that this term is entirely topological. Despite this, there will be a modification in the symplectic potential. Computing explicitly, we find 
\begin{equation} \label{eq: theta_top}
    \bm{\Theta}_{\text{top}} = - \frac{1}{4 \pi G} \delta \mathbf{A} \wedge \mathbf{F},
\end{equation}
which shows that the effect of this topological term will be identical to that of the Maxwell term (\ref{eq: Theta_M}) with a replacement of $* \mathbf{F} \rightarrow \mathbf{F}$. Putting everything together, we see that the total $U(1)$ Hamiltonian (\ref{eq: H_alpha_ren}) is modified from the form given in (\ref{eq: electric_charge}) and now reads
\begin{equation}
    H_{-1} = \frac{1}{4\pi G} \int_{\Sigma_{\infty}} \left( * \mathbf{F} + \mathbf{F} \right) = Q_e + Q_m,  
\end{equation}
i.e. the $U(1)$ charge is now the sum of the electric and magnetic charges. We note that even though only the linear combination of $Q_e + Q_m$ appears in the definition of the $U(1)$ Hamiltonian, both can be understood as conserved charges independently. Such a result follows immediately from the equation of motion $d * \mathbf{F} \approx 0$ and the Bianchi identity $d \mathbf{F} = 0$. 

As an aside, we note that (as expected) the electric and magnetic charges appear from the covariant phase space point of view as ``dual charges'', a topic of recent interest in gravitation \cite{Godazgar:2018qpq, Godazgar:2018dvh, Godazgar:2020gqd, Godazgar:2020kqd, Godazgar:2022jxm}. We will not need these more sophisticated notions as our solution of interest (\ref{eq: charged_acc_bh_metric}) does not contain non-trivial dual gravitational charges. It would be of interest to generalise the solutions to include spacetimes which do (for example those with non-trivial NUT parameter) although we leave these discussions to future work. 

\subsection{No contribution to the variational problem}

As we have modified the action via the addition of the topological term (\ref{eq: S_top}), it is natural to ask whether this may affect the variational problem analysis that was performed without the term in Section \ref{sec: variational_principle}. We note that the variation of the topological term is
\begin{equation} \label{eq: top_var}
    \delta S_{\text{top}} \approx \int d^3 x \sqrt{-g_{(0)}}  j_m^i \delta A_i^{(0)},
\end{equation}
where we have introduced the magnetic current vector field $j_{m}$, defined analogously to the electric current (\ref{eq: current}) via
\begin{equation}
j_m^i = \frac{1}{4 \pi G} \lim_{\epsilon \rightarrow 0} \left[ \frac{1}{\epsilon^3} n_{\mu} (*F)^{\mu i} \right],
\end{equation}
which can also be used to define the magnetic charge 
\begin{equation}
     Q_m =  \int_{\Sigma_{\infty}} d^2 x \, \sqrt{-g_{(0)}} j_m^t = \frac{1}{4\pi G} \int_{\Sigma_{\infty}} \mathbf{F}.
\end{equation}
Computing the values of the magnetic current explicitly for (\ref{eq: charged_acc_bh_gauge_field}), we find as the only non-zero components 
\begin{equation}
     j_m^t = \kappa \frac{g}{4 \pi G}, \qquad j_m^{\varphi} = - \frac{\alpha e}{4 K \pi G},
\end{equation}
which clearly take constant values. Combining this with $\int_{0}^{\pi} \, d\theta \sqrt{- g_{(0)}} \delta A^{(0)}_i =0$ immediately tells us that 
\begin{equation}
\delta S_{\text{top}} \approx 0 
\end{equation}
and thus the question of well-posedness is independent of whether or not one adds (\ref{eq: S_top}) to the action. We also note that the presence of the magnetic current will not affect the Ward identities (\ref{eq: Conservation_WI}), (\ref{eq: Tracelessness_WI}) as these are determined by the equations of motion which are unaffected by the addition of the topological term (\ref{eq: S_top}).

\subsection{No contribution to the first law}

Finally, we show that this topological term gives no contribution to the first law. We start by noting that this will be the case if we can show 
\begin{equation}
    \bm{k}_{\xi}^{\text{top}}  = \delta \mathbf{Q}_{\text{top}} - i_{\xi} \mathbf{\Theta}_{\text{top}}=  d \mathbf{Z} ,
\end{equation}
for some 1-form $\mathbf{Z}$, as then all contributions from the corners will cancel in the full statement of the law using the logic as argued in the paragraph below equation (\ref{eq: k}). The topological contribution to the Noether charge is
\begin{align} \label{eq: Q_top} 
\mathbf{Q}_{\text{top}}[\xi] = - \frac{1}{4 \pi G} (i_{\xi} \mathbf{A}) \mathbf{F} ,
\end{align}
which when combined with equation (\ref{eq: theta_top}) allows us to write
\begin{equation}
    \bm{k}_{\xi}^{\text{top}} =  \delta \mathbf{Q}_{\text{top}} - i_{\xi} \mathbf{\Theta}_{\text{top}} = - \frac{1}{4\pi G} \left[ (i_{\xi} \mathbf{A}) \delta \mathbf{F} + \delta \mathbf{A} \wedge i_{\xi} \mathbf{F}  \right].
\end{equation}
Using $d \mathbf{A} = \mathbf{F}$ together with $\delta_{\xi} \mathbf{A} = 0$, we can now perform elementary manipulations to reach 
\begin{equation}
     \bm{k}_{\xi}^{\text{top}} = - \frac{1}{4\pi G} d \left[ (i_{\xi} \mathbf{A}) \delta \mathbf{A} \right] \implies \mathbf{Z} = -\frac{1}{4 \pi G} (i_{\xi}\mathbf{A}) \delta \mathbf{A},
     \end{equation}
allowing us to conclude that the topological term does not contribute to the first law.

\section{Comparison with other literature} \label{sec: horizon_poly}

In this appendix we provide a careful analysis of our first law (\ref{eq: final_first_law}) with other literature, namely that of \cite{Appels:2016uha, Appels:2017xoe, Gregory:2017ogk, Anabalon:2018ydc, Anabalon:2018qfv, Cassani:2021dwa}. In the first subsection we show that our law can also be consistently derived using the ``horizon polynomial'' method of \cite{Appels:2016uha}. In the second subsection we show explicitly that ill-posedness of the variational problem results in the discrepancies between our thermodynamic lengths $\lambda_{\pm}$ given in (\ref{eq: all_physical_quantities}) and those of \cite{Cassani:2021dwa}.

\subsection{Consistency with the ``horizon polynomial" method} 

In equation (\ref{eq: final_first_law}) we wrote down the first law of accelerating black hole thermodynamics using the covariant phase space formalism, an elegant approach as this immediately allows one to identify the appearance of the conserved charges and entropy entering the law. In previous works \cite{Appels:2016uha, Appels:2017xoe, Anabalon:2018qfv, Anabalon:2018ydc, Cassani:2021dwa} a \textit{different} first law was obtained by studying the variation of the horizon polynomial 
\begin{equation} \label{eq: Horizon_perturbation}
    \delta Q(r_+) = 0,
\end{equation}
where the right hand side comes from the fact that the perturbed horizon polynomial vanishes at the perturbed horizon location.

It is natural to ask whether our law is consistent with those of \cite{Appels:2016uha, Appels:2017xoe, Anabalon:2018qfv, Anabalon:2018ydc, Cassani:2021dwa}: we have used a different method and arrived at a different result. Here we will provide a derivation of our law using equation (\ref{eq: Horizon_perturbation}), providing a useful consistency check and demonstrating that the differences in our result to those of \cite{Appels:2016uha, Appels:2017xoe, Anabalon:2018qfv, Anabalon:2018ydc, Cassani:2021dwa} are purely due to our different choices of phase space. 

We begin by recalling the definition of the horizon polynomial from (\ref{eq: metric_functions}) (with $\ell =1$)
\begin{equation} \label{eq: Horizon_poly}
    Q(r_+) = (r_+^2 -2 mr_+ + e^2 + g^2) (1- \alpha^2 r_+^2) + r_+^4 =0,
\end{equation}
an equation we will both vary and utilise on its own. Computing the variation (\ref{eq: Horizon_perturbation}) explicitly, we find 
\begin{align} \label{eq: law_midpoint}
\begin{split}
    0 & = \kappa T  \frac{2 \pi K r_+}{(1-\alpha^2 r_+^2)^2}  \delta r_+  - \frac{K}{1-\alpha^2 r_+^2} \delta m \\
   & \phantom{==} + \frac{K }{r_+(1-\alpha^2 r_+^2)} \left(e \delta e +g \delta g \right) - K \alpha r_+^2 \frac{r_+ -2m}{(1-\alpha^2 r_+^2)^2} \delta \alpha
    \end{split}
\end{align}
and recalling the definition of the Bekenstein-Hawking entropy (\ref{eq: all_physical_quantities}), we have 
\begin{equation}
    \delta S_{\text{BH}}  = \frac{2 \pi K r_+}{(1-\alpha^2 r_+^2)^2} \delta r_+ + \frac{\pi r_+^2}{1-\alpha^2 r_+^2 } \delta K + \frac{2 K \pi r_+^4 \alpha}{(1-\alpha^2 r_+^2)^2} \delta \alpha,
\end{equation}
allowing us to write the law in (\ref{eq: law_midpoint}) as 
\begin{align}
\begin{split}
     T \delta S_{\text{BH}} & = \frac{1}{\kappa(1-\alpha^2 r_+^2)} \Bigg[ \kappa T \pi r_+^2  \delta K + \frac{K r_+ ( 2 \kappa T  \pi r_+^3 +r_+^2 -2m r_+ +e^2 +g^2)}{1-\alpha^2 r_+^2} \alpha \delta \alpha \\
     & \phantom{aaaaaaaaaaaaa} + K \delta m - \frac{K}{r_+} \left( e \delta e + g \delta g \right) \Bigg].
\end{split}
\end{align}
All that remains now is to write the right hand side as $\delta \mathcal{M} -\Phi_e \delta Q_e - \Phi_m \delta Q_m + \lambda_- \delta \mu_- + \lambda_+ \delta \mu_+$. This calculation is straightforward but a little tedious and requires use of equation (\ref{eq: all_physical_quantities}) for the definitions of $\mathcal{M}, \Phi_e, Q_e, \Phi_m, Q_m, \lambda_{\pm}, \mu_{\pm}$. After some algebraic manipulation we arrive at 
\begin{equation}
    T \delta S_{\text{BH}} = \delta \mathcal{M} -\Phi_e \delta Q_e - \Phi_m \delta Q_m + \lambda_- \delta \mu_- + \lambda_+ \delta \mu_+ + C,
\end{equation}
where 
\begin{align}
\begin{split}
  C & =  \frac{Km}{\kappa^2} (1 - \alpha^2 \Xi ) \delta \kappa  \\
  & \phantom{aaa} +  \bigg( \frac{r_+}{\kappa} - \frac{r_+^3}{(r_+^2\alpha^2-1) \kappa} - \frac{m(3+2\alpha^2)}{2\kappa} - (e^2 +g^2) \frac{mr_+ \alpha^4-1}{r\kappa}  \Bigg) \delta K \\
  & \phantom{aaa} +  \bigg( m \left[\alpha ^2 r_+^2-1\right] \left[-\alpha ^2 \left(e^2+g^2\right)+r_+^2 \left(\alpha ^2+\alpha ^4 \left(e^2+g^2\right)+4\right)-1\right] \\
  & \phantom{aaaaaaaaa} +2 r_+ \left[-\alpha ^2 r_+^2 \left(e^2+g^2+r_+^2\right)+e^2+g^2+r_+^4+r_+^2\right] \Bigg) \frac{\alpha K  \delta \alpha  }{(1-\alpha^2 r_+^2)^2 \kappa}
  \end{split}
\end{align}
and now utilising the phase space relationship (\ref{eq: variational_constraint_general}) required for well-posedness of the variational problem we can simplify this to 
\begin{equation}
    C =  Q(r_+) \frac{\delta K(1- \alpha^2 r_+^2) +2 \alpha  \delta \alpha  K r_+^2}{\kappa  r_+ \left(\alpha ^2 r_+^2-1\right)^2} = 0,
\end{equation}
where we used (\ref{eq: Horizon_poly}) in the final step, thus demonstrating equivalence between the covariant phase space and horizon polynomial approaches.

\subsection{Differences in thermodynamic lengths} \label{sec: differences}

We will now demonstrate the reason for the differences in our ``thermodynamic length'' parameters $\lambda_{\pm}$ to those of the previous literature \cite{Appels:2016uha, Appels:2017xoe, Gregory:2017ogk, Anabalon:2018ydc, Anabalon:2018qfv, Cassani:2021dwa}. We compare our law with that of \cite{Cassani:2021dwa}, as that is the case with the largest number of non-trivial parameters and in particular has $g \neq 0$. Their general first law is given in equation (3.46) of that work and in order to compare with ours we set
\begin{equation}
    J = 0, \qquad \ell = 1, \qquad  \delta \ell = 0, \qquad \delta G = 0,
\end{equation}
reducing equation (3.46) of \cite{Cassani:2021dwa} down to 
\begin{equation} \label{eq: first_law_cassani}
    \delta \mathcal{M}_{\text{hol}} = T \delta S_{\text{BH}} + \Phi_e \delta Q_e + \Phi_m \delta Q_m - \tilde{\lambda}_- \delta \mu_- -\tilde{\lambda}_+ \delta \mu_+,
\end{equation}
where all quantities above are equivalent to those defined in (\ref{eq: all_physical_quantities}), \textit{except} for the thermodynamic lengths, which are given by 
\begin{equation}
    \tilde{\lambda}_{\pm} = \frac{r_+}{\kappa (1\pm \alpha r_+)} - \frac{m}{\kappa \Xi} \mp \frac{\alpha \Xi}{\kappa}. 
\end{equation}
We also note that we have explicitly put the holographic mass $\mathcal{M}_{\text{hol}}$ as defined in (\ref{eq: holographic_mass}) in the first law when written as in (\ref{eq: first_law_cassani}). This is a crucial feature of this law as we shall now demonstrate.

In order to compare with \cite{Cassani:2021dwa} we remove the requirement of well-posedness (\ref{eq: variational_constraint_general}) and choose $\kappa$ as given in (\ref{eq: kappa_old}). Using equation (\ref{eq: H_ren_mass}) together with the definitions of the string tensions (\ref{eq: cosmic_string_tensions}) we find that the difference between the variations of the masses is now 
\begin{align} \label{eq: mass_differences}
     \delta H^{\text{ren}}_{\partial_t} - \delta \mathcal{M}_{\text{hol}}=  \frac{m(2\alpha^2 \Xi -1)}{\kappa \Xi} \delta( \mu_+ + \mu_-),
\end{align}
which we note is no longer zero.\footnote{Unless one fixes the overall deficit $\delta (\mu_+ + \mu_-) = 0$, e.g. in the case of a spindle.} This means that the first law is sensitive to the choice of mass that appears, for example we can rewrite (\ref{eq: first_law_cassani}) with the Wald Hamiltonian as
\begin{equation} \label{eq: first_law_cassani_wald}
    \delta H^{\text{ren}}_{\partial_t} = T \delta S_{\text{BH}} + \Phi_e \delta Q_e + \Phi_m \delta Q_m - \lambda_- \delta \mu_- -\lambda_+ \delta \mu_+,
\end{equation}
where the right hand side of (\ref{eq: mass_differences}) has been absorbed into the thermodynamic lengths, which are now in agreement with our expression for $\lambda_{\pm}$ given in (\ref{eq: all_physical_quantities}). 

This calculation demonstrates explicitly that the removal of a well-posed variational problem results in an ambiguous first law which changes form depending on the choice of mass charge that one uses. If the variational problem is well-posed then both notions of mass are equivalent and one finds a single consistent first law as given in (\ref{eq: final_first_law}). We thus strongly advocate well-posedness as a requirement in order to establish equality between various notions of mass and to obtain consistent thermodynamic relations.



\bibliographystyle{JHEP}
\bibliography{draft.bib}

\end{document}